# The hunt for research data: Development of an open-source workflow for tracking institutionally-affiliated research data publications

Bryan M. Gee[1,*]

[1]University of Texas Libraries, The University of Texas at Austin, TX
*corresponding author*: bryan.gee@austin.utexas.edu
*ORCID*: 0000-0003-4517-3290

## Abstract

The ability to find data is central to the FAIR principles underlying research data stewardship. As with the ability to reuse data, efforts to ensure and enhance findability have historically focused on discoverability of data by other researchers, but there is a growing recognition of the importance of stewarding data in a fashion that makes them FAIR for a wide range of potential reusers and stakeholders. Research institutions are one such stakeholder and have a range of motivations for discovering data, specifically those affiliated with a focal institution, from facilitating compliance with funder provisions to gathering data to inform research data services. However, many research datasets and repositories are not optimized for institutional discovery (e.g., not recording or standardizing affiliation metadata), which creates downstream obstacles to workflows designed for theoretically comprehensive discovery and to metadata-conscious data generators.

  Here I describe an open-source workflow for institutional tracking of research datasets at the University of Texas at Austin. This workflow comprises a multi-faceted approach that utilizes multiple open application programming interfaces (APIs) in order to address some of the common challenges to institutional discovery, such as variation in whether affiliation metadata are recorded or made public, and if so, how metadata are standardized, structured, and recorded. It is presently able to retrieve more than 4,000 affiliated datasets across nearly 70 distinct platforms, including objects without DOIs and objects without affiliation metadata. However, there remain major gaps that stem from suboptimal practices of both researchers and data repositories, many of which were identified in previous studies and which persist despite significant investment in efforts to standardize and elevate the quality of datasets and their metadata.

# 1. Introduction

Data discoverability (Findability) is a core principle of open scholarship and the FAIR principles (Wilkinson et al., 2016). A data deposit that is optimally findable can be discovered by both humans and machines and without prior knowledge of its existence (e.g., datasets are often known to exist *a priori* through an avenue such as an associated article). Additionally, data should be findable not only by specialist researchers or by researchers in general but also by other entities who may be interested in finding and reusing data in various capacities (e.g., Wallis et al., 2006; Tenopir et al., 2011; Borgman, 2012; Gregory, 2020; Gregory et al., 2020; Sun et al., 2024). Ensuring broad findability is intrinsically related to maximizing reuse of data, which can occur outside of the research ecosystem, even when data were generated within that context. For example, the Generalist Repository Ecosystem Initiative (GREI) recognizes four potential use cases of research data (Staller et al., 2023; Van Gulick et al., 2024): (1) a researcher who wants to share data; (2) a researcher who wants to find data; (3) a funder who wants to monitor compliance and to track impact; and (4) an institution who wants to monitor compliance and to track impact. There are nuances within and in addition to these use cases (e.g., different units within an institution; an instructor seeking data for pedagogical, rather than research, purposes), but they provide a starting foundation for thinking more broadly about possible motivations for data discovery.

Academic institutions have long monitored research outputs; historically, the focus has been on articles, books, and commercializable products (e.g., patents), but with the growth of the open scholarship movement and associated implementations like data sharing mandates, there is growing interest in tracking a wider range of outputs (e.g., research data publications, open-source research software). Efforts to discover datasets published by affiliated researchers have been undertaken at a number of institutions using various conceptual approaches and practical solutions; most of those that are published in some form have been performed by librarians and digital scholars (e.g., Barsky et al., 2016; Lafia & Kuhn, 2018; Mannheimer et al., 2021; Sheridan et al., 2021; Briney, 2023; Alawiye & Kirsch, 2024; Dellureficio & Pokrzywa, 2024; Johnston et al., 2024; Wang, 2024; Lostner & Krzton, 2025; Warner, 2025).

These efforts have identified just as many obstacles to discovery as they have identified potential solutions and workflows. Johnston et al. (2024), a multi-institutional study conducted as part of the Realities of Academic Data Sharing (RADS) initiative, provided one of the most detailed summaries of many of the existing infrastructural and cultural challenges around efficient, comprehensive discovery of institutional outputs. These include: (1) variation in workflows, such as the quantity and quality of metadata crosswalked to DataCite or Crossref; (2) support for, and utilization of, persistent identifiers such as ORCID (Open Researcher and Contributor ID) and ROR (Research Organization Registry); (3) selection of required (minimum) metadata fields; (4) quality control measures such as curation; and (5) linkages (or lack thereof) to related research objects such as articles, preprints, and software.

Commercial solutions that have either been designed specifically for research data (e.g., Elsevier's Data Monitor) or that have been co-opted and expanded from a previously more limited scope of journal articles (e.g., Clarivate's Web of Science) are available and widely used by colleges and universities. However, these proprietary solutions come with a number of philosophical and logistical shortcomings that run counter to many principles of the open scholarship movement. These include subscription fees (reflecting broader commercialization of scholarly research); revenue-driven decision-making (e.g., how to develop and whether to maintain products); semi-arbitrary platform restrictions (e.g., a daily cap on the number of retrieval requests); limited interoperability between other systems and workflows (e.g., ability to share out retrieved outputs); closed-source processes that make it more difficult to assess accuracy and completeness of retrieved records; and biases in inclusion/exclusion of content. Many of these have been well-documented for platforms originally designed for journal articles (e.g., Vieira & Gomes, 2009; Franceschini et al., 2016; Mongeon & Paul-Hus, 2016; Vera-Baceta et al., 2019; Zhu & Liu, 2020; Pranckutė, 2021; Visser et al., 2021) and exist for datasets as well (e.g., Benjelloun et al., 2020; Chapman et al., 2020; Sostek et al., 2024).

     Especially for datasets, even the most extensively developed solutions are intrinsically limited by upstream infrastructure and processes in data repositories (e.g., utilized metadata schemas, quality and crosswalking of metadata; Wu et al., 2019; Gregory et al., 2020; Löffler et al., 2021) and behavior of researchers around repositories, such as use of non-repositories avenues for sharing data that lack long-term preservation mechanisms (e.g., GitHub, cloud storage, personal websites, journal-hosted supplemental information [SI], and "upon personal request"; Santos et al., 2005; Krawczyk & Reuben, 2012; Kenyon et al., 2016; Alter & Gonzalez, 2018; Dutra do Reis et al., 2021; Tedersoo et al., 2021; Hussey, 2023; Sharma et al., 2024). Especially with the perpetual challenge of distributing limited library budgets to many areas of need, finding opportunities to avoid subscription dependency is important for ensuring library sustainability. The recent announcement of the shuttering of Elsevier's Data Monitor (Elsevier, 2025), a closed-source solution that cannot be continued by another entity, further underscores the risks associated with subscription-based commercial solutions.

     Development and adoption of open-source solutions offers an alternative. Open-source software (OSS) is widely deployed in a variety of academic, non-academic, and mixed-use scenarios, ranging from QGIS (QGIS, 2025), a freely available open-source alternative to the commercial ArcGIS Pro software for managing geospatial data, to the Dataverse Project (Crosas, 2011), an open-source platform for developing data repositories that is used by more than 125 institutions at present. OSS is not without potential shortcomings, including the frequent need to develop them from scratch and to maintain them without significant funding or staffing support, but there are also significant advantages. These include transparency in the principles and mechanisms underlying a process; opportunity for broader community engagement with the product (which can mitigate limitations of developing software without commercial support); flexibility to tailor the processes for specific use cases and edge cases; avoidance of subscription dependencies; and the ability for others to freely access, reuse, and repurpose the software (helping to ensure long-term viability if the original developer[s] can no longer continue to maintain it). In addition to the technical benefits of such an approach, open-source solutions also exemplify the best practices and principles around research data and software management that many digital scholars promote to researchers: that metadata, data, and software should be made maximally FAIR.

This paper describes the development of an open-source, Python-based workflow for tracking research data publications at the University of Texas at Austin (hereafter, 'UT Austin'). Although the development was done in the specific context of this university, the workflow can be easily reconfigured to search for research data published by researchers at any other institution or series of institutions. The primary workflow is a broad-scope retrieval through the DataCite API and accounts for variation in which metadata field(s) affiliation metadata can be recorded and variation in how affiliations can be constructed when not standardized using ROR identifiers. It is supplemented by a series of more targeted secondary processes that utilize other APIs to focus on specific repositories or specific modalities of data sharing that lack affiliation metadata altogether, that only contain poorly-structured affiliation metadata, or that do not mint DOIs. These supplementary processes result in a composite workflow that has higher coverage potential than previous OSS workflows (e.g., Johnston et al., 2024). The static version of the code used for this version of the preprint is archived on Zenodo (Gee, 2025a), while the codebase is actively developed on GitHub (https://github.com/utlibraries/research-data-discovery) under a permissive reuse license. It is recognized that Python is not the preferred language for all potential reusers who may nonetheless be interested in the conceptual framework for the workflow given the multi-faceted retrieval process. This paper thus focuses on providing a conceptual overview of the framework underpinning this workflow, demonstrating some of its analytical power, and discussing challenges (and their potential or realized solutions) encountered during its development in order to increase reuse potential.

## 2. Methods

It must be emphasized that this workflow was developed and refined through extensive trial-and-error and exploratory work, not always with a fully-clarified vision, that led to iterative refinement of the processes. Although it is presented in a coherent fashion by virtue of norms for presentation of scholarly papers, it was not seamlessly written into its current form from the beginning. The patchwork nature of development of this workflow, particularly components related to unexpected metadata and repository nuances, is also a reflection of my combined experience as an active scientific researcher who has published data in both generalist and specialist repositories (e.g., Gee & Sidor, 2024; Gee et al., 2025) and as a former data curator at Dryad who has experience in data curation and metadata workflows, in addition to my current role providing support for UT Austin's institutional data repository, the Texas Data Repository (hereafter 'TDR'). This positionality is noted mainly because it contextualizes certain specific real-world case studies or examples discussed later as well as the overall nature of how this workflow was developed.

### 2.1. Overview of the conceptual approach

#### 2.1.1. Search approach

There are two primary avenues to institutional discovery. The first avenue is to rely on author names, searching for records that match to entries in a registry of affiliated researchers (e.g., Mannheimer et al., 2021, for Montana State University). Limitations to this approach include:

(1) the ability to efficiently create and update databases of researchers likely requires collaboration with other internal units (e.g., human resources) and potential use and management of controlled-access data; (2) difficulty efficiently disambiguating records authored by researchers with common names given the variable adoption of ORCIDs; (3) difficulty tracking records across all operational tiers within a research unit (i.e. not all outputs include a permanent employee like a faculty member as a coauthor); and (4) difficulty efficiently disambiguating work that may have been done at another institution (i.e. researchers may publish work from their previous institution with that affiliation while at the present institution of interest, or they may publish work conducted at the present institution after they have moved to another institution).

The second avenue is to rely on affiliation metadata, searching for a particular institutional name or set of names (e.g., Johnston et al., 2024, for RADS Phase I institutions, hereafter 'the RADS study'). Limitations to this approach include: (1) variation in how institution names may be constructed (e.g., 'UT Austin' vs. 'University of Texas at Austin'); (2) variation in whether the parent institution is included in the affiliation of a dataset (i.e. a researcher may list their affiliation only as an subsidiary institute or school and omit the parent institution); (3) sensitivity and precision of search systems (e.g., an API may operate on exact match only and therefore not detect an affiliation that includes unexpected formatting or granular information); and (4) variation in whether affiliation metadata are recorded (affiliation is not a required field in Crossref or DataCite), and if so, whether they are crosswalked into a standardized schema like DataCite. Based on our assessment of the options, a decision was made to use the second method.

A third approach that is not presently feasible for most institutions but worth a brief mention is the examination (manual or automated) of data availability statements based on a curated catalog of institutional scholarly outputs (theses & dissertations, articles, preprints, books) from which information on data sharing can be mined (e.g., the approach of Briney 2023, 2024 for Caltech). This approach requires extensive resources to be dedicated to obtaining a comprehensive set of records and then curating records for which dataset connections are not present in the metadata or for which data are not shared through links (e.g., supplemental information, repositories without link-backed PIDs). This approach may become more feasible in the future if institutions take an active approach to facilitating compliance with public access policies of U.S. funding agencies (e.g., requiring their researchers to store green OA versions of all institutionally-affiliated articles in an IR).

### 2.1.2. Utilized data sources

The primary source of data for this affiliation-based workflow is the DataCite REST API, which is freely and publicly available without requiring an API token. Unlike some institutions that may have previously minted DOIs for institutionally-managed repositories through Crossref (e.g., Johnston et al., 2024), UT Austin does not oversee any data repositories that do so. Like the RADS study, early exploration of purportedly affiliated records minted through Crossref returned primarily objects that appear to be incorrectly classified as 'datasets' (e.g., peer reviews in Faculty Opinions). The lack of support for Boolean operators in the Crossref REST API creates an additional hurdle for public institutions like UT Austin whose full name contains many generic terms (contrasted with institutions like Cornell or Stanford, for example) that lead to reduced accuracy of returned results (i.e. many 'false positives').

The RADS study made use of a previous version of the Crossref public data file, which is one way to circumvent the API's limitations, but I elected not to use this dataset here because of concerns over whether its rapidly growing size makes it an unsustainable resource for an average user. The file has practically quadrupled to over 200 GB in the few years since its initial release (though the recently released 2025 version is slightly smaller than the 2024 version; Crossref, 2025), and it may be, or will soon reach, a size that is intractable for an average user in terms of the storage required to maintain it, the time required to download it, and the skills and resources necessary to process it. The calculus on whether this dataset is viable and how to maintain it locally if desired will depend on the frequency of dataset retrieval and the objectives for doing so — a single snapshot like the RADS study versus monthly reporting, as we would like to do at UT Austin, will invoke different needs and costs, especially given the frequency at which metadata records are edited in Crossref and DataCite (e.g., Hendricks et al., 2020; Strecker, 2024). Similar considerations apply to the DataCite public data file (over 350 GB in uncompressed form for the 2024 version; DataCite, 2024). Queries for datasets through the Crossref API are discussed further below.

A large number of other APIs were explored or incorporated into different parts of the retrieval process. In addition to the DataCite and Crossref APIs, I also queried the publicly accessible REST APIs for Dataverse (specifically the TDR endpoint), Dryad, Figshare, and Zenodo (Dataverse and Zenodo require users to create an account and to register a free API key) for information about datasets. I also made use of the publicly accessible OpenAlex REST API as part of secondary processes to identify affiliated datasets through linked articles.

### 2.1.3. Organization

The workflow utilized here is Python-based (in contrast to the R-based approach of Johnston et al., 2024; code available in Mohr & Narlock, 2024) and makes use of the following actively maintained and widely used modules: *datetime* (Python standard library); *io* (Python standard library); *json* (Python standard library); *math* (Python standard library); *numpy* (Oliphant, 2006); *os* (Python standard library); *pandas* (McKinney, 2011); *requests* (Reitz, 2025); *selenium* (Goucher et al., 2025); *urllib* (Python standard library); and *xml* (Python standard library). The workflow was developed in Python version 3.12.5. The decision to use Python was motivated largely by existing workflows of the UT Austin Research Data Services team, which, in turn, partially reflect observations that Python is vastly more popular than many other programming languages (e.g., R), both in the general coding community (GitHub, 2024) and at UT Austin (internal workshop data).

The codebase comprises multiple components divided among several scripts (Fig. 1). The primary workflow is considered the 'core' part of the retrieval process and is responsible for retrieving the majority of records from the DataCite REST API. Secondary processes contained within the same script, which can be toggled on/off via Boolean variables, perform targeted searches that are customized for specific use cases or repositories (e.g., a workaround to identifying affiliated Figshare deposits that do not themselves record affiliation metadata). Finally, there are accessory scripts that were used for proof-of-concept (e.g., a script that queries the DataCite API using the ROR identifier in order to quantitatively compare the efficacy of different search methods), summary work (e.g., retrieving summarized metadata from the DataCite API for select publishers), and data visualization. Some scripts have been designed exclusively for this preprint but are provided in the software deposit (Gee, 2025a).

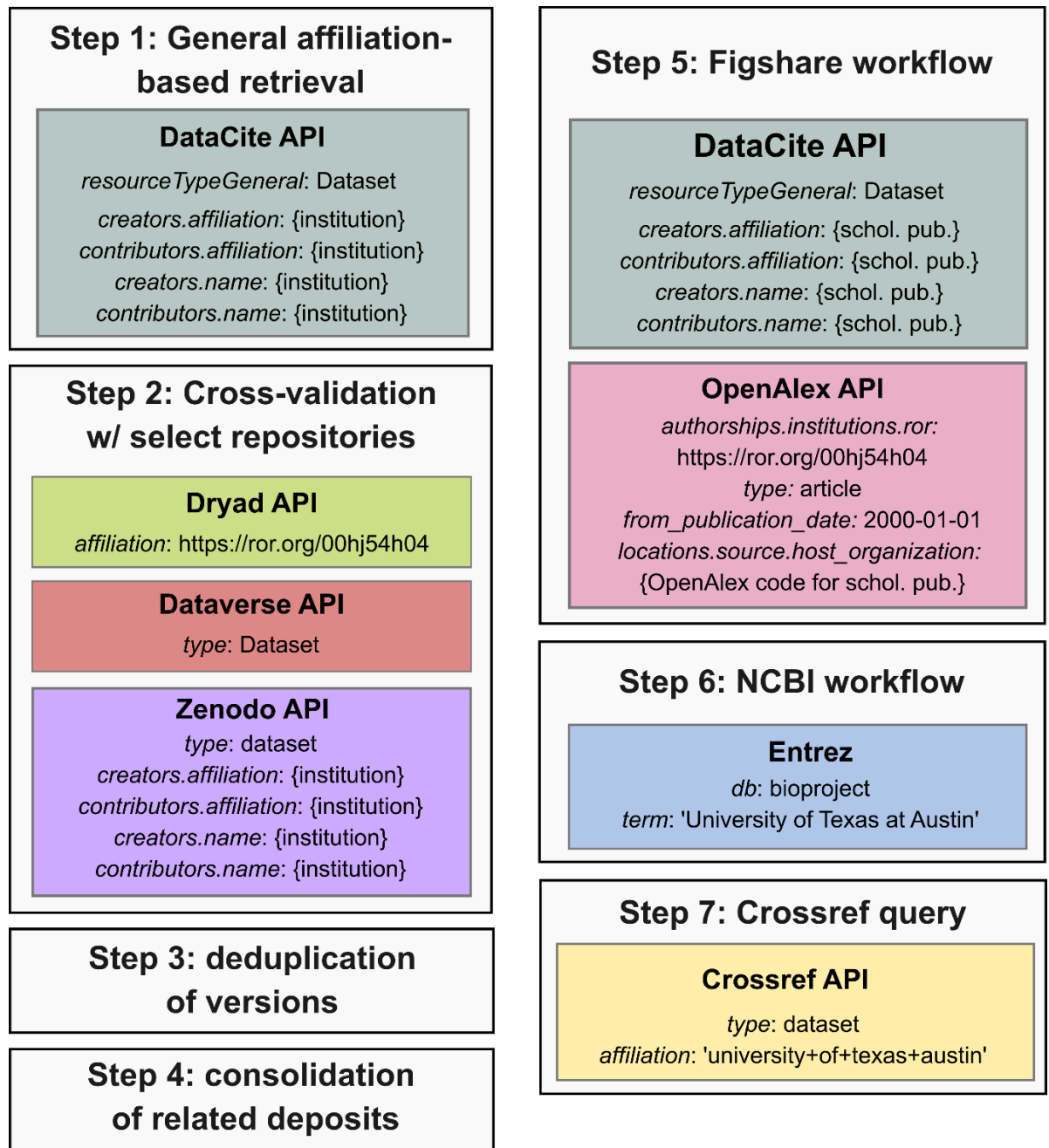

**Figure 1. Schematic overview of the current dataset discovery workflow.** '{institution}' represents a list of permutations of the institutional name. 'schol. pub.' ('scholarly publisher') represents a list of Figshare publisher partners who mint Figshare DOIs through DataCite in this workflow and '{OpenAlex code for schol. pub}' is an internal code in OpenAlex that is assigned to each scholarly publisher (this is a list of focal publisher partners specifically). No affiliation parameter is included for the Dataverse API query because it is querying specifically the TDR instance rather than all Dataverse installations.

## 2.2. Primary search workflow

### 2.2.1. Development of workflow

The primary limitation of the affiliation-based search process is the lack of standardization of how affiliations are entered. Johnston et al. (2024) commented on how the relative lack of adoption of ROR identifiers hinders the use of this PID for institutional discovery. To demonstrate the effects of limited ROR adoption for UT Austin specifically, Table 1 summarizes the results from a ROR-based query in the DataCite API, specifying UT Austin's ROR identifier in the *creators.affiliation.affiliationIdentifier* field and objects labeled as 'dataset' in the *type.resourceGeneral* field, which returned only a few hundred results skewed towards Dryad (an early adopter of ROR; Gould & Lowenberg, 2019).

**Table 1. Listing of retrieved dataset counts using a ROR-based query for affiliated research datasets in the DataCite API.** A total of 893 DOIs across 18 repositories were initially retrieved. Cleaning steps are the same as those of the primary workflow that is described later in this section; cleaning reduced the count to 571 DOIs. All repositories with fewer than five entries in the initial retrieval are grouped together into the final row. Data as of July 1, 2025.

| Repository | Initial count | Post-cleaning count |
|---|---|---|
| Dryad | 404 | 404 |
| Zenodo | 257 | 109 |
| Figshare | 181 | 15 |
| Mendeley Data | 15 | 7 |
| NERC EDS UK Polar Data Centre | 8 | 8 |
| Science Data Bank | 7 | 7 |
| NOAA NCEI | 5 | 5 |
| Other repositories | 16 | 16 |
| **TOTAL** | **893** | **571** |

I then modified the query to search for a string with the official formulation of UT Austin, "The University of Texas at Austin," rather than the ROR identifier; this process retrieved slightly more than 1,000 results (Table 2), some of which are functional duplicates (when a repository mints two or more DOIs for the same deposit, like with Figshare or Zenodo). Crucially, the ROR-based query did not retrieve any deposits from our institutional data repository, the Texas Data Repository (TDR), a Dataverse-based repository that has more than 1,300 UT Austin-affiliated datasets with DataCite DOIs (without capacity for including ROR identifiers in the affiliation metadata until early 2025). The single-affiliation-based query only recovered 46 TDR datasets, collectively indicating that a more expansive search process was necessary.

**Table 2. Listing of retrieved dataset counts using a single-affiliation-based query for affiliated research datasets in the DataCite API.** The string utilized was 'The University of Texas at Austin,' the official permutation of the institution, and was queried only in the *creators.affiliation* field. A total of 1,210 DOIs across 34 repositories were initially retrieved. Figshare and Figshare+ are grouped together here. Cleaning steps are the same as those of the primary workflow that is described later in this section; cleaning reduced the count to 769 DOIs. All repositories with fewer than five entries in the initial retrieval are grouped together into the final row. Data as of July 1, 2025.

| Repository | Initial count | Post-cleaning count |
|---|---|---|
| Dryad | 404 | 404 |
| Zenodo | 414 | 177 |
| Figshare | 193 | 21 |
| Harvard Dataverse | 49 | 49 |
| Texas Data Repository | 46 | 46 |
| ICPSR | 22 | 9 |
| IEEE DataPort | 16 | 2 |
| Earth System Grid Federation | 13 | 13 |
| Digital Porous Media Portal | 10 | 10 |
| PhysioNet | 6 | 2 |
| NSF Arctic Data Center | 6 | 5 |
| Other repositories | 33 | 31 |
| **TOTAL** | 1,210 | 769 |

In order to identify the metadata attributes that led to an incomplete retrieval, I ran several cross-validation processes against specific repositories' APIs. To do so, I queried the DataCite API for UT Austin-affiliated deposits, with the added specification for a specific repository in the *publisher* field (e.g., Zenodo), made an equivalent affiliation-based query to the repository's API, and then cross-referenced the DOIs returned from each API to identify deposits found in one query but not the other. This was done for Dryad, TDR (functional equivalent to Harvard Dataverse), and Zenodo. Originally, I intended to perform this process for all of the GREI repositories (except for Vivli, which is specifically for clinical data and not a true generalist). However, investigation of other repositories' APIs revealed that this is not possible for Figshare, which usually only records affiliation metadata for institutional members, and for OSF, which only records affiliation metadata for institutional members. Mendeley Data originally only crosswalked affiliation metadata for institutional members, but began crosswalking all affiliation metadata around the initiation of GREI (based on personal examination of UT Austin data). For Mendeley Data deposits without affiliation metadata in DataCite, it may be possible to retrieve this metadata from the Mendeley Data API — affiliation metadata is recorded for non-member institutions on dataset landing pages — but this API requires a request for access to be made (instead suggesting the use of OAI-PMH). Without any details on how requests for access are considered, I did not pursue this option, as it was unclear whether this API would be accessible to others interested in reproducing or adapting this workflow (including others at UT Austin).

This cross-validation process efficiently identified dozens to hundreds of affiliated datasets in these repositories that were returned by a repository-specific API but not by the DataCite API (or occasionally the inverse). I manually examined random datasets to discern why they were not recovered by the DataCite API and identified a wide range of metadata discrepancies with three major explanators: (1) a non-ROR-linked affiliation entry can vary widely (e.g., 'University of Texas, Austin' is a permutation of UT Austin that appears in some datasets; Fig. 2); (2) affiliation metadata may be more granular than the university level (e.g., Durkan & Warburton, 2023, which includes department and postal code) or may be multi-institutional (e.g., Lichtenberg et al., 2017); and 3) affiliation metadata may appear in fields other than *creators.affiliation*; for example, more than a hundred records were only detected via listing of the institution as an entity equivalent to an individual researcher (Fig. 3).

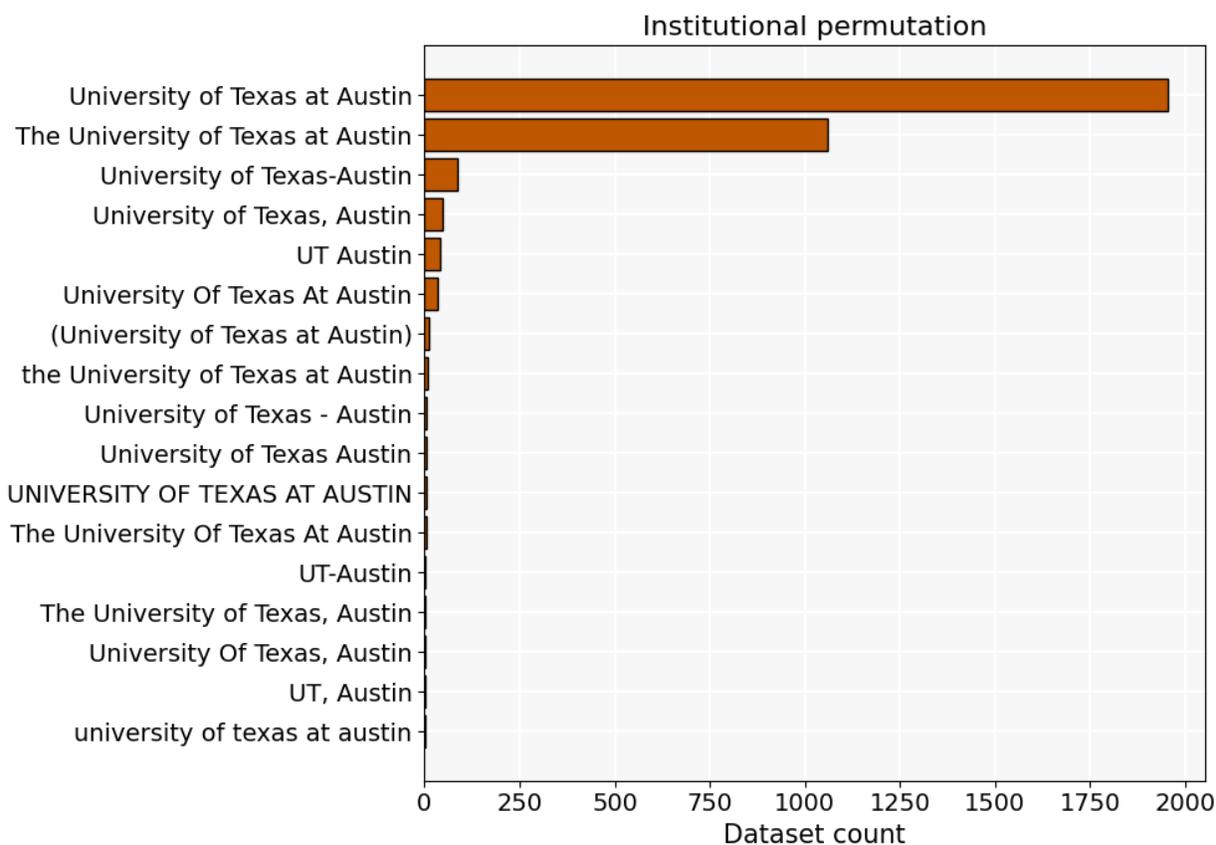

**Figure 2. Comparison of the frequency of different permutations of "The University of Texas at Austin" among datasets retrieved through the DataCite API.** Each permutation shown here occurs at least once in the corpus of datasets retrieved through the DataCite general query or cross-validation with specific repositories (*n* = 2,888); some permutations are included in the search query but were not detected. In some instances, the permutation shown is part of a more granular affiliation (e.g., with departmental information) and was only detected through the cross-validation process; only the institutional part is shown here. The workflow does not currently look for misspellings of all of these permutations since there is a limit to how many can be queried at once in DataCite's REST API, but these are known to exist. For NCBI, additional permutations were detected, but because these do not need to be explicitly queried (a single institutional query is sufficient for detecting many permutations), they are not shown here. Data as of July 1, 2025.

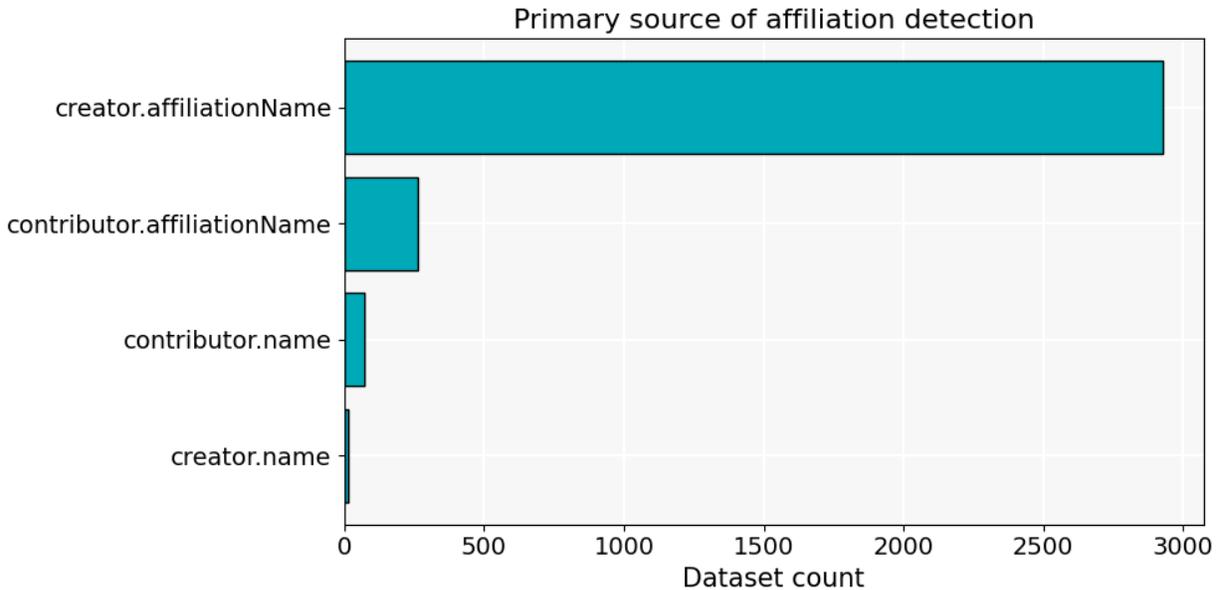

**Figure 3. Comparison of the frequency of DataCite fields in which a permutation of UT Austin was detected.** Each dataset is only classified into one bin and thus only counted once. The classification scheme uses the hierarchy of '*creator.affiliationName*,' '*contributor.affiliationName*,' '*creator.name*,' and '*contributor.name*.' In other words, if an affiliation is detected in multiple fields for one dataset (e.g., *creator.affiliationName* and *contributor.affiliationName*), it is categorized as whichever field comes first in the hierarchy (e.g., *creator.affiliationName*). The hierarchy is based on the relative intuitiveness of searching a given field (e.g., I would expect to find affiliation in an affiliation field more often than as a creator/author). The data depicted represent all datasets retrieved in the general affiliation-based DataCite query and the cross-validation with specific repositories' APIs (*n* = 2,888). Data as of July 1, 2025.

These metadata attributes are not surprising, but they have not typically been accounted for in previous work either. The first point explains why previous queries in testing of this workflow did not discover deposits in TDR because the institution name was previously crosswalked in parentheses into the DataCite metadata: '(University of Texas at Austin)'; following the identification of this formatting through this workflow, TDR updated the crosswalk to remove parentheses (early 2025). Because the DataCite API requires exact string matches for many fields, other minor deviations from the official permutation, such exclusion of a leading 'The', cause datasets to not be detected without being individually queried. Johnston et al. (2024) utilized the *rdatacite* R package (Chamberlain & Kramer, 2023), which permits non-exact searching that is useful in some instances (e.g., detection of datasets where a researcher enters more granular affiliation metadata than just the institution), but for certain institutions, it is intrinsically sensitive to inclusion of 'false positive' datasets that contain a different affiliation of similar construction. For example, manual examination of data from the RADS study (Mohr & Narlock, 2024) revealed some datasets authored by a researcher for 'Duke Kunshan University' that were labeled as being affiliated with 'Duke University' and some authored by a researcher at 'West Virginia Institute of Technology' that were labeled as being affiliated with 'Virginia Tech'.

The current version of the primary workflow thus involves a query that looks for one of over 35 permutations of UT Austin's name (Fig. 2), not all of which presently occur in a DataCite record, across four metadata fields: *creators.affiliation*; *contributors.affiliation*; *creators.name*; and *contributors.name* (Fig. 3). It searches for both datasets and software, but the focus of this preprint and the majority of records are datasets, and most of the results presented are exclusively for datasets (rationale provided in the next section).

Although the cross-validation process was originally intended only to identify additional permutations to be incorporated into this filter, non-exact affiliation matching in specific repositories' APIs also identified affiliated deposits that contain one institutional permutation as part of a more detailed string (e.g., 'Department of Integrative Biology, University of Texas at Austin') and that cannot be efficiently retrieved in the DataCite API due to their specificity and number (there is a limit to how many affiliations can be queried in one DataCite call). This part of the primary workflow takes the list of affiliated DOIs that were only identified from the specific repositories' APIs and retrieves their DataCite entries individually in order to add them to the output.

### 2.2.2. What about code?

Johnston et al. (2024) analyzed both institutionally affiliated datasets and software. This workflow mainly focuses on datasets for a few reasons. Firstly, there is more cultural and infrastructural emphasis on open data sharing than there is on open code sharing, which leads to lagging rates of code sharing compared to data sharing, even when researchers on a given project generate both outputs (e.g., Culina et al., 2020; Hamilton et al., 2022, 2023; Kambouris et al., 2024; Maitner et al., 2024; Sharma et al., 2024). These gaps are evident in everything from job responsibilities (e.g., many research data librarians focus primarily or exclusively on data management and sharing, not on software, although experience with common software used to analyze data is a frequent qualification) to journal and funder mandates (e.g., most U.S. federal agencies developed policies around data sharing but not code sharing).

The UT Libraries are certainly interested in open code sharing as one of the partners of the UT Open Source Program Office (UT-OSPO), but we are also aware of three realities: (1) that fewer researchers will utilize or generate code than will utilize or generate data; (2) standards and policies around code publishing, specifically deposition in PID-minting repositories rather than common software development platforms like GitHub that do not mint PIDs (among other preservation deficiencies), lag well behind those of data publishing; and (3) researchers may not always differentiate 'software' from 'data' and thus either publish their code alongside data as part of mixed-media deposits or publish standalone code in deposits labeled as 'dataset' (or another object type in the schema). Separate but related work to identify UT Austin-affiliated GitHub repositories has identified over 35,000 GitHub repositories managed by more than 1,600 unique accounts (Gee & Shensky, 2025), indicating that even an approximation of research software outputs is unlikely to be attainable through DOI-minting platforms anytime soon. Early testing also revealed an extreme domination of the affiliated software DOIs by Zenodo, so I felt that it was best to focus on analyzing and comparing research dataset publications, with future work intended to more comprehensively explore various forms of code sharing. A short section on software is presented in the Results. As a note, the primary workflow can be modified to search for additional object types in the DataCite schema by creating a Boolean expression for the variable that defines what resource types to look for.

## 2.3. Secondary search workflows

Additional sources of UT Austin-affiliated datasets are known to exist but cannot be as efficiently retrieved or require custom workflows; these are detailed in this section as secondary workflows. As previously noted, these are largely contained in the same script as the primary workflow but are not enabled by default and can be toggled on/off, as they will return fewer results than the broad DataCite API query.

### 2.3.1. Crossref API

Depending largely on whether an institution has previously used Crossref to mint DOIs for one or more of its managed data repositories, Crossref may yield a significant number of additional datasets or only a small number. As previously noted, UT Austin has not previously used Crossref for this purpose, and because of the many common words in the institution's name, the non-Boolean Crossref API would be expected to (and did) (1) return a large number of putatively affiliated deposits; and (2) return predominantly unaffiliated deposits (e.g., ones associated with other UT system campuses). For this reason, the Crossref API call is externalized as a separate script because I do not anticipate querying this source on a frequent basis for UT Austin. A re-user could insert it into the script containing the primary workflow however. This secondary workflow is relatively straightforward, using 'university+of+texas+austin' as the affiliation query and searching for objects labeled as 'dataset' ('software' is not an object type in Crossref). It returns an output that is formatted to be concatenated with the DataCite API output.

### 2.3.2 Other sources of datasets

Beyond datasets that are discovered through the DataCite or Crossref APIs through affiliation metadata, there are two main categories of deposits that may represent affiliated research datasets but that cannot be discovered in the same fashion: (1) deposits that have DOIs minted through DataCite or Crossref but that lack affiliation metadata; and (2) deposits that either use another PID system that is not centrally searchable (e.g., ARKs, handles) or that have no PID at all (e.g., GitHub, personal websites, most supplemental information) but that do contain affiliation metadata in some internally standardized form. This section discusses the development of two secondary workflows that begin to target these groups of semi-invisible deposits.

#### 2.3.2.1. Affiliated datasets with DOIs but without affiliation metadata: the Figshare case study

As previously noted, some repositories, including well-known generalists, do not record affiliation metadata or only do so in narrow instances (e.g., only for paying member institutions in the case of OSF). Figshare is another repository that typically only records affiliation metadata for institutional members, but it differs from OSF in some respects that make it possible to discover certain affiliated Figshare deposits, even when an institution is not a member. Both UT Austin and all but one of the RADS institutions are not institutional members, but both Johnston et al. (2024) and the primary workflow presented here do recover affiliated Figshare deposits. Investigation of individual deposits both provides insight into potential explanators and verifies the suspicion that the Figshare record of affiliated datasets will be relatively incomplete for any given institution.

Firstly, the number of recovered Figshare deposits is much lower than what would be expected based on the relative volume of total deposits in Figshare compared to other generalist repositories. For example, in both the RADS study and this workflow, the number of discovered affiliated Figshare deposits is lower than that of either Dryad or Zenodo deposits, but the latter two, particularly Dryad (<65,000 datasets; Dryad, 2025) have far fewer total objects than Figshare (>8 million objects; Figshare, 2025). Unless researchers have a particular aversion to Figshare, I would not expect to see a significant difference in the proportional representation of generalist repositories among high research activity institutions.

Secondly, inspection of Figshare deposits that list affiliation metadata in DataCite and that were recovered for UT Austin (subsequent to development of the primary workflow) and for the RADS institutions (prior to development of the primary workflow; Johnston et al., 2024) reveals that virtually all deposits with Figshare listed as the *publisher* are connected to Springer Nature articles (Table 3). This was determined by targeting the *relatedIdentifier* field from these deposits for any identifier where the dataset is listed as 'I*sSupplementTo*' and then querying the Crossref API to gain more information about these articles. This observation is noteworthy for two reasons. Springer Nature is part of Holtzbrinck Publishing Group, as is Digital Science, the latter of which funds Figshare. Additionally, Springer Nature is one of many large publishers who partner with Figshare to use a workflow in which materials that are uploaded as 'supplemental information' (or equivalent) in partner journals' manuscript submission systems are automatically published on Figshare upon article publication (there are some deviations in which authors provide additional metadata that is crosswalked; e.g., Springer Nature, n.d.). Examples of other notable current partners include Frontiers, PLOS, and Taylor & Francis (Figshare, n.d.b). This integration is largely invisible to authors and even to editors; authors usually do not have to have a Figshare account or log into the platform. My knowledge of this process is based on first-hand experience as a researcher – I personally "published" eight Figshare deposits, mostly through Taylor & Francis titles (e.g., Gee et al., 2019d, for Gee et al., 2019c), before discovering that this is how supplemental information is hosted by the publisher. Given the partnership with non-Springer Nature titles, I would expect to see non-Springer Nature titles represented among the discovered affiliated Figshare datasets.

**Table 3. Comparison of publisher volume of distinct articles associated with RADS Figshare deposits.** The RADS dataset (Mohr & Narlock, 2024) was filtered for objects labeled as 'dataset' and with 'figshare' listed as the *publisher* (omitting any deposits mediated through a non-Springer Nature partner; *n* = 2,276) and then cleaned in the same fashion as in the workflow of this preprint, resulting

in 1,040 DOIs after removal of versions. Among these 1,040 DOIs, there are 268 unique article DOIs listed as being supplemented by a Figshare deposit, across 86 unique journal titles. All but one of them is published by Springer Nature.

| Journal | Publisher | Count |
| --- | --- | --- |
| *BMC Plant Biology* | Springer Nature | 27 |
| *BMC Genomics* | Springer Nature | 22 |
| *Genome Biology* | Springer Nature | 14 |
| *BMC Biology* | Springer Nature | 11 |
| *BMC Bioinformatics* | Springer Nature | 10 |
| *BMC Cancer* | Springer Nature | 7 |
| *Genome Medicine* | Springer Nature | 7 |
| *Molecular Cancer* | Springer Nature | 7 |
| *Journal of Experimental & Clinical Cancer Research* | Springer Nature | 6 |
| *Journal of Neuroinflammation* | Springer Nature | 6 |
| *Journal of Translational Medicine* | Springer Nature | 6 |
| *Microbiome* | Springer Nature | 6 |
| *BMC Medicine* | Springer Nature | 5 |
| *Cell & Bioscience* | Springer Nature | 5 |
| *Journal of Hematology & Oncology* | Springer Nature | 5 |
| *Stem Cell Research & Therapy* | Springer Nature | 5 |
| All other Springer Nature titles (*n* = 68) | Springer Nature | 118 |
| *CABI Agriculture and Bioscience* | CABI Publishing | 1 |
| **TOTAL** | | **268** |

Thirdly, and related to the previous remark, both this workflow and Johnston et al. (2024) do retrieve affiliated Figshare deposits that are linked to non-Springer Nature publishers, but they are indexed differently than those linked to Springer Nature articles. For Springer Nature-linked datasets, the *publisher* is listed as 'figshare' in the DataCite schema, and the connection to Springer Nature is only revealed by obtaining additional information about the individual DOI linked as a related identifier in the metadata (though Springer Nature DOIs are sometimes recognizable). For other publishers who partner with Figshare and who mint mediated DOIs through DataCite (not all of them do; e.g., PLOS), the publisher in the DataCite schema is listed as the scholarly publisher itself (e.g., a Figshare deposit mediated through a Taylor & Francis title will list 'Taylor & Francis' as the *publisher* in DataCite) which more clearly establishes the connection. This accounts for what may be an initially surprising discovery of datasets that are listed as being published by a scholarly publisher who does not actually publish data.

Even accounting for these nuances in the listed publisher, the volume of discovered affiliated Figshare deposits is still clearly well below the expected volume. In part, this relates to mediated Figshare deposits that had DOIs minted through Crossref (e.g., PLOS titles), as these deposits have much lower-quality metadata (affiliation, related identifiers) than mediated deposits with DataCite DOIs. The Crossref DOIs often can only be reliably or efficiently connected to an article through the construction of the DOIs, which often appends '.s00*' to the end of the article DOI where '*' is an integer (e.g.,

'*10.1371/journal.pone.0297637.s001*' is the first supplemental file for the *PLOS ONE* article with the DOI '*10.1371/journal.pone.0297637*'; Woolley et al., 2024a, 2024b). Observations of a pervasive lack of recording of affiliation in these mediated deposits is arguably the most vexing obstacle. However, Figshare is one of the largest volume repositories, and it thus seems important to find a way to capture at least a better estimate of the number of datasets of interest in this repository.

In order to begin developing a custom workflow for trying to identify potential Figshare deposits, I first surveyed the list of publisher/journal partners listed on Figshare's website and recorded select metadata based on random examination of multiple recent deposits for each entity (summary file included in Gee, 2025b). It should be noted that the list of Figshare partners has certainly changed over time (e.g., SAGE was listed as a partner in early 2025 and is no longer listed; Figshare, n.d.a); this can introduce some additional challenges because datasets mediated through a previous partner will presumably still retain the same metadata following a change in status of the partnership. The recorded metadata include which DOI registry the deposit is minted through; how the DOI is constructed (i.e. the core prefix); who the listed publisher of the dataset is; and what the client or provider ID is. As significant heterogeneity was identified among such deposits, three different approaches to identifying affiliated Figshare deposits that do not themselves contain affiliation metadata are listed below.

*2.3.2.1.1. DataCite-minted objects*

The mediated Figshare process results in variation in metadata across nearly every metadata field that is regularly crosswalked to DataCite. Two attributes are relied upon for this process. The first attribute was previously noted: for all publisher partners who mint DOIs through DataCite, other than Springer Nature, the listed *publisher* is not 'figshare' but rather the scholarly publisher (e.g., 'Taylor & Francis'). Because these publishers do not maintain their own data repositories, and their articles are minted through Crossref, it can be reasonably inferred that any DataCite-minted DOI that lists a scholarly publisher partner as the *publisher* is a mediated Figshare deposit. This provides an effective filter to programmatically identify the full set of mediated Figshare deposits for one or several publishers through the DataCite API. The second attribute is that because of the linkage to a manuscript, these mediated Figshare deposits usually record the relationship to the published article as a relationship (*relationType*: '*IsSupplementTo*') with the associated article DOI. This workflow is diagrammed in Figure 4.

This secondary workflow queries the DataCite API for all deposits with a *type.resourceGeneral* classification of 'dataset' and a *publisher* listing of a publisher partner that is known to mint DOIs through DataCite (e.g., Taylor & Francis). Note that mediated deposits with 'data' proper may be classified as another object type, but the total volume of mediated deposits is so high that it was deemed more efficient to only focus on those labeled as 'dataset' at this stage. Likewise, not all objects labeled as 'dataset' contain data, but assessing the object classification of deposits is a cross-repository challenge and not directly addressed here. Most of these deposits do not contain any affiliation metadata, but most contain the DOI of their respective linked article. Next, I queried an API with article metadata to retrieve articles published by the same publisher (e.g., Taylor & Francis) and that are specifically authored or coauthored by a UT Austin researcher. This workflow uses OpenAlex, which uses ROR identifiers to standardize metadata ingested from Crossref, making for a more efficient API call, but Crossref's API or public data file, or another API (e.g., Unpaywall), could alternatively be used. The two dataframes are then cross-matched using the article DOIs, which will link mediated Figshare deposits without affiliation metadata to articles authored by UT Austin researchers (Fig. 4).

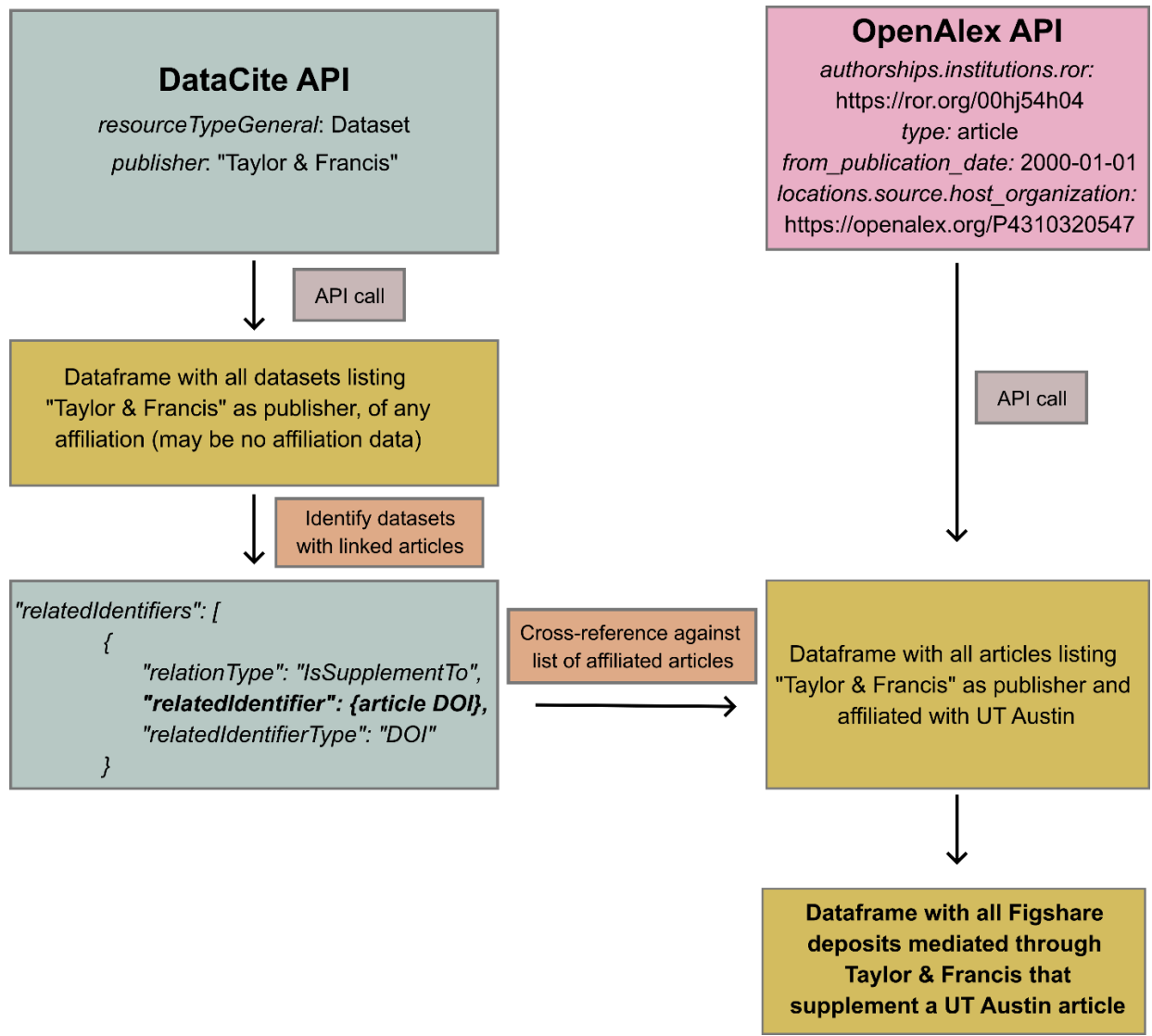

**Figure 4. Schematic diagram showing the secondary workflow to discover affiliated Figshare deposits that were mediated through a publisher partner and that usually lack affiliation metadata.** The code listed in the *locations.source.host_organization* field for the OpenAlex API is the OpenAlex code for Taylor & Francis. In the full workflow, a list of publisher partners' names and OpenAlex codes' is looped in a single API call.

*2.3.2.1.2. Crossref-minted objects*

Compared to mediated Figshare deposits minted through DataCite, those minted through Crossref have comparatively poor metadata. They do not formally indicate the article they are related to or author information. There is also an extremely large number of them because each file uploaded as supplemental information receives its own DOI, even though it is highly unlikely that any researcher would manually create a separate deposit for each file associated with one manuscript. This can occur when minting DOIs through DataCite but does not always and appears to be publisher-specific. For example, a search for Crossref objects published by PLOS and labeled as 'component' (how Figshare deposits mediated through PLOS are labeled) returns more than 4 million records. This makes a process like that described above for mediated Figshare deposits with DataCite DOIs impractical given the time necessary just to retrieve all such objects, but it would be far more tractable with the public data file.

Two different approaches have been tested to programmatically retrieve these Figshare deposits with Crossref DOIs. The first is to identify a list of affiliated articles published by a partner like PLOS and then to test the existence of a hypothetical SI file that has a '.s00*' or '.t00*' suffix added to it. When tested with PLOS (about 875 UT Austin-affiliated articles at present), this process proved to be rather time-consuming because it requires individual queries to verify the existence of each hypothetical DOI and because it will not return useful metadata if it does exist (e.g., if this verification was done with the Crossref API). Some articles have more than 20 SI files, and because of the lack of metadata about an object's true contents, all of them have to be retrieved for collective assessment (e.g., SI files 1-19 could be accessory non-data files and file S20 could be a raw data file). This approach is included in the codebase, but it is generally discouraged for use and is provided mainly to demonstrate that this was explored.

The second approach, again tested with affiliated PLOS articles, is to use text-scraping (through the *beautifulsoup4* module; Richardson, 2025) to target the HTML blocks that contain information about SI files. For PLOS articles, this provides more information than the Crossref API because the landing page contains information like the title of the deposit, which includes a value from a semi-controlled vocabulary (e.g., Figure, Table, Text), a description of the file, and the file format (e.g., DOCX, XLSX, PDF). Additional fields could either be scraped from the page or retrieved from the Crossref API for the article (e.g., author information). This additional metadata would be important if one wanted to go beyond merely noting the number of existing DOI-backed deposits (e.g., to use file formats to assess whether something is 'data' proper). This process was more time-efficient than the previously described one, but it is intrinsically sensitive to any future modifications to the HTML code, and it would require a custom approach for each publisher partner's distinct HTML formatting. Additionally, not all of the publisher partners who use Crossref for mediated Figshare deposits provide the same level of metadata for SI content in the web version of articles. In early stages of testing, there was also some variation in how metadata were entered between PLOS titles, and ostensibly this would exist within other publishers as well.

The final approach, which also only works with PLOS, is quick but coarse-grained and involves the use of the PLOS Open Science Indicators (OSI) dataset (Public Library of Science, 2024). This dataset, which is on version 9 and under ongoing development, involved analysis of the entire corpus of PLOS articles for data and software sharing, which includes a categorization of where data were shared. The secondary workflow involving this dataset retrieves a list of affiliated PLOS articles and then filters the PLOS OSI dataset for any affiliated article that has data shared in part or in whole through SI, which is in actuality the mediated Figshare process. This method assumes that the PLOS algorithms for identifying 'data' are philosophically aligned with those of the investigator and cannot directly provide additional information on the count, nature (e.g., file formats), or DOIs of these mediated deposits. It also relies on periodic versioning of the core dataset to be useful in the long-term. Given these limitations, this approach is unlikely to be widely used, but it proved useful in early testing and exploration, runs extremely quickly (<60 seconds), and might be sufficient for certain use cases (e.g., a quick approximation), so it is provided as an accessory script.

2.3.2.2. Discovering deposits without DOIs that contain affiliation metadata: the NCBI case study

The counterpart to discovering DOI-backed deposits without affiliation metadata is discovering non-DOI-backed deposits with affiliation metadata. Many repositories do not mint DOIs for their deposits, either at all or by default, and arguably the majority of research data have been shared without a DOI (e.g., Briney, 2023). This can occur not only through SI but also in well-established repositories. For example, NCBI (National Center for Biotechnology Information) repositories like GenBank issue non-digital PIDs like accession and project numbers. These are unique and persistent but are not digital in the same way as DOIs. In other instances, a data deposit can be made public ("published") without being issued a DOI, even though one can be issued. Probably the best-known example is Open Science Framework (OSF). The platform provides a simple means to mint a DOI for a project on the project's landing page, but this still requires a researcher to be aware that the project URL is not a DOI and to manually select this option. Based on personal experience with researchers, because minting a DOI is not required for making a deposit public, researchers may not mint a DOI (or do not realize that the landing page link is not a DOI) and instead publish a non-DOI link in a data sharing statement (OSF project links look similar to DOIs and are incorporated into DOIs if minted; example project link: '*https://osf.io/t659s/*'; and corresponding DOI: '*https://doi.org/10.17605/osf.io/t659s*').

      Deposits that do not automatically have, or cannot have, DOIs minted for them are not intrinsically bound to the DataCite metadata schema, but this does not preclude these deposits from recording affiliation metadata in an internally standardized fashion. Therefore, it may be possible to discover institutionally-affiliated deposits without a DOI within a specific repository, even if it cannot be done across many repositories through a single source/query like the DataCite or Crossref APIs. Whether targeting a given repository is worth the time necessary to investigate the infrastructure for batch queries is dependent on the estimated or known volume of deposits. This section describes efforts to programmatically retrieve deposits linked to UT Austin from NCBI, which is known to hold nearly 1,000 affiliated BioProjects (closest approximation of a 'dataset').

      NCBI was identified early on as a major gap in retrieval because of the longstanding centralization of bioinformatics data across NCBI repositories, none of which mint DOIs. It will likely be a significant source of data for any institution with measurable NIH funding (and/or

NSF funding, to a varying degree). For NCBI, affiliation metadata is recorded and searchable in the web interface, but it lacks a REST API equivalent to that of DataCite or Crossref. Instead, NCBI has a centralized system called Entrez, which can be accessed through its in-house command-line tool called E-Utilities or through a third-party library/module like *eutils* (Stevenson et al., 2019) or *biopython* (Cock et al., 2009). However, there are rate limits and a request to only make large requests during off-peak hours (9 PM-5 AM Eastern), which may be suboptimal for some users. Two approaches are described here, both of which are included as a secondary workflow.

The first approach makes use of *biopython*, which directly queries Entrez and generates an XML file that is then converted to a dataframe for concatenation with the other outputs. This is a relatively straightforward scripted approach that is mainly dependent on the maintenance of the module and the size of the query not exceeding the rate limiting. The second approach uses the Selenium Webdriver, a tool typically used for web development and testing in which actions within a web browser can be scripted (e.g., to click on a button to download a file). The second approach scripts retrieval of BioProjects by simulating a user manually downloading the full record of search results. It opens a browser window to a link specifying BioProjects and with a search of "University+of+Texas+at+Austin" across all fields and then mimics a user manually clicking on a drop-down button and selecting a series of options to download an XML file with metadata on each BioProject. As with the first approach, this script converts the XML to a dataframe for concatenation. This process is faster than the use of *biopython* because loading the page and scripting the steps is not intensive, but similar to text-mining, it is dependent on stability of the HTML, which is reasonable for a federally managed repository. Reusers should familiarize themselves with NCBI guidelines on use of scripting calls to NCBI servers (National Center for Biotechnology Information, 2025). Given that neither process takes more than about 30 seconds, the greatest advantage of the Webdriver approach is that it is likely to be more transferable if there are other high-volume platforms where a similar approach would work (e.g., single-click download for all queried results), even if the HTML structure is different. It is less likely that a third-party library/module like *biopython* will have functionality to make queries across multiple independent repositories, and the time necessary to learn how to use a new package and/or API for each additional repository could be more costly than making slight modifications to a Selenium-based workflow.

### 2.3.3. Cleaning steps

An additional significant component of this workflow is the cleaning and filtering of retrieved deposits. For example, a well-known source of variation among data repositories is in PID assignment, both with respect to the type of PID (e.g., DOI, ARK, handle) that is assigned and the granularity of PID assignment (e.g., dataset vs. individual file; single PID for all versions of a dataset vs. different PID for each version). Repositories that use Dataverse software may mint a DOI for each individual file, in addition to each dataset, while others like Zenodo mint DOIs for each version of a dataset. This section describes the current measures to attempt to standardize the scale of dataset records between repositories.

Similar to Johnston et al. (2024), this workflow removes all file-level DOIs of Dataverse installations, which are easily recognizable as they contain an additional suffix appended to the end of the DOI for the dataset and which record this relationship in metadata. Note that as recently as 2019, Dryad also minted DOIs for individual files (by adding '/*' to the end of the dataset DOI, where '*' is a sequential integer). This possibility is accounted for in this workflow, but it did not actually apply to any UT Austin-affiliated Dryad records and may similarly be irrelevant for other institutions that have never been Dryad members. As I did not manually examine every single deposit retrieved in the workflow, it is possible that other repositories may have a different system for file-level DOI assignment that are not immediately apparent from cursory examination of DOI structure.

As with Johnston et al. (2024), this workflow handles the multi-DOI approach of Zenodo, which mints at least two DOIs for each dataset: a 'parent' that always resolves to the most recent version and a 'child' for each version (there will be two DOIs, even if only one version of a dataset exists). In most cases, the 'parent' and first 'child' DOIs will be sequential, but subsequent 'child' DOIs (second version onward) will not have any obvious relationship from visual examination of the DOIs. Only one DOI (the 'parent') from each 'lineage' is retained here.

Zenodo's DOI versioning system differs from other generalist repositories that append 'v*' to the end of a DOI, where '*' is an integer (e.g., v1 is for version 1), to create DOIs for new versions. For example, Figshare, ICPSR, and Mendeley Data mint DOIs in the same 'parent-child' fashion as Zenodo, but their use of the 'v*' system makes the DOIs' relationship more easily discernible than Zenodo. In this system, the 'parent' lacks any 'v*' ending. It should be noted that some repositories' practices around DOI assignment for versions have evolved over time (e.g., very old Figshare deposits use a single DOI for all versions). In some instances, not all 'child' DOIs are recovered, even if the 'parent' is recovered. The reason for this is most likely progressive evolution of affiliation metadata. For example, Van de Vuurst (2021) is a Figshare dataset with eight published versions, but only the parent and versions 6, 7, and 8 were retrieved from the DataCite API by Johnston et al. (2024) based on the dataset of Mohr & Narlock (2024). Manual examination of the DataCite metadata of versions 1 through 5 that were not retrieved reveals that no affiliation metadata is recorded, suggesting that the dataset was initially published without this information and then had it added (Virginia Tech, a RADS institution and Figshare member, appears in later versions). It seems most likely that the addition of affiliation metadata is related to a repository increasing its capacity for recording this information, changing its processes for crosswalking this information (compared to an author choosing to update their deposit after not initially recording this information, which is always hypothetically possible but which seems less likely to occur based on general metadata entry behaviors of authors), or re-curating deposits. Among the initial DOI set retrieved for UT Austin, I also observed one instance in which when the parent Figshare DOI is not listed with affiliation metadata, but specific versions' DOIs, including the most recent version that that parent should resolve to, were listed with this metadata (Cai et al., 2023).

This workflow departs from Johnston et al. (2024) in removing all but the first or 'parent' version of a dataset for repositories with the 'v*' DOI versioning system, which mirrors the approach taken for Zenodo. When comparing the results for UT Austin to those of RADS institutions (Johnston et al., 2024), this step accounts for significant differences in counts between certain repositories. Future work could explore which aspects of a dataset are most often changed in a new version (e.g., metadata vs. data files) and the degree to which they change (e.g., adding one new file vs. replacing all files), but in my experience, versioning tends to occur at a relatively limited level that does not warrant treatment of two versions as conceptually separate datasets for purposes like comparing total number of deposits between repositories.

   This workflow also departs from Johnston et al. (2024) in further treatment of Figshare deposits, in light of the nature of how mediated deposits are created. Beyond the decision of how to handle the 'parent-child' DOI system, one of the sources of variation in the mediated Figshare process is whether multiple files uploaded for a single manuscript are published under a single DOI or are split out into one DOI for each file (there may also be an encapsulating DOI labeled as a 'collection' in the DataCite schema that "houses" many DOIs in the latter instance). When a Figshare DOI is given to each file, the granularity is like that of Dataverse installations, which inflates the number of objects labeled as 'dataset' if not accounted for ('file' is not an object type in DataCite). Although there are certainly scenarios where materials for one project may need to be split into multiple deposits (either within one repository or across multiple; e.g., for mixed licensing or different authorship), examples that I encountered for this Figshare process appear to be a byproduct of automation rather than active choice and do not present obvious metadata differences that warrant their separation. I have never seen an instance in which a researcher decided to manually create a separate DOI-backed deposit for each supplemental table in a generalist repository, for example, but this is a frequent occurrence through the Figshare mediated process. Therefore, the second step of cleaning for Figshare deposits targets the *relatedIdentifier* field, specifically the related article being supplemented, and combines all datasets supplementing the same article DOI into a single entry. This consolidation process is also applied to deposits that list a publisher partner as the dataset publisher (e.g., where 'Taylor & Francis' is listed as the publisher) instead of 'figshare,' although it should be noted that not all publisher partners over-split files for one manuscript. This step significantly further reduces the number of retained Figshare deposits and widens the disparity between counts for the RADS institutions and UT Austin.

   The final aspect that departs from Johnston et al. (2024) is in optional further processing of Dataverse deposits. An observation made in TDR is that some authors create multiple deposits, each labeled as 'dataset' in the Dataverse parlance (a DOI-backed container), that are all for a single manuscript/article and that are nested under a single 'dataverse' (a non-DOI-backed container). Examples in TDR include five datasets (code, reconstruction parameters, system parameters, results, measurements) for one study on zebrafish embryos (Yang, 2024a, 2024b, 2024c, 2024d, 2024e); and four datasets (two types of raw data, experimental conditions, README) for one study on mechano-lysis in blood clots (Rausch, 2025a, 2025b, 2025c, 2025d). In most instances, authors seem to be separating materials only based on format or an internal delimiter (e.g., samples, specimens), rather than along different authorship lists or with different licenses/terms of use for these separate-but-related deposits. In contrast to the mediated Figshare process, they are usually not splitting each file into its own deposit, and instead there is some logical basis for separation. As with the Figshare process, this practice partially inflates the number of Dataverse 'datasets' and is probably the result of the availability of the 'dataverse' container, whereas there is rarely an equivalent in other repositories. This step in the workflow treats objects with the same publication date (down to the day), list of creators (names and affiliations), and rights as partial duplicates and consolidates them into a single entry. It is possible to loop each dataset through the Dataverse Native API endpoint to retrieve direct information on the encapsulating dataverse, but this is a more time-intensive step if there are many deposits, and some Dataverse installations have fairly restrictive rate limiting. Researchers may also use the same dataverse for multiple projects, each with multiple datasets, so simply being in the same dataverse does not mean that multiple projects are all directly related. The resultant counts from this Dataverse deduplication process can be drastically lower than the original, post-file-removal count: if this is applied to the UT Austin datasets, of the 1,328 dataset-level DOIs in TDR that are affiliated, only 868 (65%) are retained after this step. As the over-splitting of these deposits is different than either the mediated Figshare process or the assignment of file-level DOIs in Dataverse, it is considered a more optional step. For the Results section, the data presented do not incorporate this additional consolidation step.

There are a few other processes involved with cleaning at the level of smaller-volume repositories to individual datasets; most of these are likely to be specific to UT Austin. For the Environmental Molecular Science Laboratory, it appears that individual but related samples are deposited with separate DOIs in this platform, as there are sets of datasets with identical metadata (authors, publication date, etc.), separate DOIs, and the same type of data in each deposit (e.g., 15 deposits titled 'Data for EMSL Project 47414 from March 2020'). These were deduplicated based on a combination of metadata fields (dataset title, first author, *relationType*, *relatedIdentifier*, *containerIdentifier*). For DesignSafe, affiliation metadata is suboptimally entered in the *contributor.name* field, rather than in an affiliation field, and it is not apparent which affiliation corresponds to which author(s). Because the repository is hosted at the Texas Advanced Computing Center (TACC) at UT Austin, all datasets list 'University of Texas at Austin' as one of the 'contributors' regardless of whether a UT Austin researcher is listed as a PI or co-PI for the deposit (verifiable on the landing page). Cursory examination of random deposits indicated that where a UT Austin researcher is listed as a PI/co-PI, the affiliation will be specifically 'University of Texas at Austin (utexas.edu)', although not all deposits were checked to confirm that this is a consistent and presently utilized crosswalk. This attribute was used to remove DesignSafe entries without this particular form of the affiliation. Finally, listed repository names were corrected and standardized. For example, some Zenodo datasets do not list Zenodo as the *publisher*; as this is a freeform editable field in Zenodo (in contrast to many other repositories). As a result, "rogue values" can be entered in this field. There are valid reasons for entering a value other than the hosting repository in certain instances, but for the purposes of this work, any deposit with 'zenodo' in the DOI string was changed to list Zenodo as the publisher. Reusers of this workflow will need to manually check their own outputs, as additional outliers will likely be unique to an institution (the ability for authors to edit the *publisher* field exists in other platforms; the Global Biodiversity Information Facility [GBIF] is another example). Repositories that have multiple permutations of their name listed (e.g., 'Texas Data Repository' vs. 'Texas Research Data Repository') also had their names standardized.

### 2.3.4. Metadata assessments

Different organizations or institutional units will have different objectives for usage of data about discovered datasets. For example, some institutions' primary aim in discovering datasets may be to centralize them in a data catalog (e.g., Sheridan et al., 2021; Creel, 2025; Warner, 2025). At UT Austin, one of our objectives is metadata assessments to understand attributes like trends of our institutional data repository (TDR) compared to other repositories; large data publishing; prevalence of different file formats; and the frequency with which code is being published within objects labeled as 'datasets'; and in turn, to use that to inform the development of our research data services. This section describes some of the current metadata assessment functionality. It must be noted that in DataCite and Crossref, the relatively low bar for minimum metadata results in marked heterogeneity in the presence and form of some information, such as whether deposit size is crosswalked and if so, whether it is as a list of sizes for each file or a total sum. As a result, with a few exceptions for required fields (e.g., anything assessing dataset titles), only a subset of the retrieved datasets can be analyzed for these assessments.

#### 2.3.4.1. Object classification

For DOI-backed deposits, one of the major challenges is whether the resource type classification is accurate. Johnston et al. (2024) demonstrated this point in their discovery of a large number of objects labeled as 'dataset' in the Faculty Opinions LTD platform that are not datasets and are only labeled as such due to the lack of a more appropriate label in the schema. I also encountered this problem particularly with Crossref records. Presumably, this workflow and others are also failing to retrieve DOI-backed deposits that contain, in part or in whole, data but that are labeled as another object type (e.g., the Crossref 'component' of PLOS supplemental information files). Objects can be mislabeled for a variety of reasons, such as mixed-object deposits (e.g., data and software); conceptual differences between and among researchers and repositories over the idea of 'data' (e.g., qualitative data like transcripts of historical texts may be less likely to be thought of as data if they were not generated specifically for research; Guy et al., 2013; Gualandi et al., 2023); and researchers' lack of familiarity with the schemas when using repositories that require them to select an object type. Especially in un-curated generalist platforms that accept a wide range of non-data objects (e.g., Figshare, OSF, Zenodo), researchers' ability to select from the full array of object types undoubtedly affects metadata quality.

     A combination of personal experience outside of this project and discoveries within the context of this project suggests that object mislabeling is a serious obstacle to institutional discovery, both with respect to objects that are not discovered and with respect to objects that are discovered but that are misclassified. One workaround is the development of models to identity and characterize mentions of research data or software in text of articles (e.g., Howison & Bullard, 2016; Zhao et al., 2018; Du et al., 2021; Istrate et al., 2022, 2024; Schindler et al., 2022; Pan et al., 2023; Druskat et al., 2024). However, there are various challenges to executing this approach at the institutional level: (1) whether a comprehensive set of university-affiliated articles can be identified; (2) whether the text of those articles can be scraped (some subset or even majority is assured to not be available for

paywalled articles); and (3) whether researchers' concepts of terms like 'dataset' and 'software' align with those of librarians or other digital scholars who are attempting this discovery process (e.g., some researchers consider anything that is not in the main text to be 'data'). A similar approach based on metadata within a dataset (e.g., description field) is likely to be inhibited by sparse or uninformative metadata population (e.g., verbatim abstract of the associated article).

File formats offer another potential avenue, provided that this information is crosswalked to DataCite or Crossref. However, there are limitations, mainly that researchers often publish data in suboptimal formats (e.g., tabular data in PDF format) or mixed-media combinations (e.g., image and tabular data in Microsoft Word files), so it is difficult to establish absolute rules that have a low error rate. Filename information is unreliable, assuming it is crosswalked, because of conceptual and operational variation in terms like 'appendix,' 'dataset,' 'software,' and 'supplemental information.' It cannot be assumed that any object labeled as 'supplemental' in the filename is not 'data'; for example, supplemental figures in many natural history disciplines comprise specimen photographs taken in conformity with longstanding conventions and are usually not considered 'creative' outputs (e.g., Egloff et al., 2017). Nonetheless, certain relatively absolute rules can be defined. For example, some formats are unambiguously not 'data' in this study's context, such as files in programming languages like Python (.py, .ipynb) or R (.r). This workflow can thus make some assessments of whether objects include code in addition to data (i.e. mixed-media deposits) or are only code (and should be labeled as 'software').

### 2.3.4.2. Authorship position

It is virtually assured for almost any institution that not all datasets with an affiliated researcher were created, curated, or published by that researcher and thus do not necessarily reflect any active decision-making or other contributions by an affiliated person. Especially given variability in how researchers determine dataset authorship (e.g., same author list as associated manuscript vs. only data uploader) and the lack of an "author contributions" section (with standardized roles like the CRediT taxonomy), the role of an affiliated researcher in a given dataset publication is often unclear. However, inferring this role is important when trying to assess patterns in the data; for example, one repository might appear to be preferred by university researchers based on comparison of datasets with a second repository, but this could alternatively be the result of an unaffiliated collaborator's frequent preference for the first repository, and the affiliated researcher(s) may have played no role in picking the repository. Based on current conventions in which the first author (of any output) tends to have contributed the most, and the last author is often the PI or group lead, this workflow uses the first and last authors' affiliations to classify datasets based on whether the focal affiliation occurs in the first author position, the last author position, both positions, neither position, or there is only one author. This is treated as one coarse estimator of datasets that were actively managed by an affiliated researcher.

### 2.3.4.3. Other metadata assessments

Several other assessments are currently enabled in the workflows. Extracting licensing/rights information 'as-is' from the DataCite API can offer useful insights into another area in which researchers are often engaging in suboptimal practice (e.g., attempting to use legal Attribution clauses in Creative Commons licenses to force the non-legal community best practice of citation). This also relates directly to the prevalence of software publication in deposits labeled as 'datasets.' Information on file formats can give some insight into another area of researcher behavior: whether researchers are publishing data in maximally accessible formats (e.g., an open standard format like CSV [comma-separated values] vs. a proprietary format like XLSX [Microsoft Excel]). Total deposit size information is useful for understanding the patterns of 'large data' publishing, such as which platforms seem to be preferred and how common are datasets above a given size threshold. Finally, analyzing dataset titles to determine how descriptive they are using lists of nondescript words like grammatical articles can provide insight into where education can be targeted to improve basic metadata provision for datasets. The metadata assessment components of this workflow remain relatively underdeveloped and are intended for further work as part of this project and others (e.g., frequency of ORCID usage), although there are significant limitations to what can be gleaned from a central API like that of DataCite compared to a specific repository's API (e.g., filename information).

## 3. Results

All data summarized here are based on records collected on July 1, 2025. Because the various scripts were run sequentially over several hours, not concurrently (and thus in theory, a script run at the end might include a few additional datasets not retrieved in an earlier script). The full corpus of data across the various datasets produced from this workflow is thus best viewed as containing all records indexed with a relevant date through June 30, 2025. Through the combination of workflows, more than 4,000 distinct data deposits were identified as being affiliated with UT Austin. They are summarized by their workflow source (e.g., DataCite API vs. Crossref API) in Table 4 and by total deposit count in Table 5.

**Table 4. Summary results of the five sources of affiliated research datasets that were utilized in this workflow.** 'DataCite API' represents the primary workflow that makes a broad query for objects labeled as 'dataset' and with one of many permutations of UT Austin. 'Specific repositories' APIs' represents additional datasets that are indexed in DataCite, labeled as 'dataset,' and contain, in part, some permutation of UT Austin but that were not detected in the general query and only discovered through cross-validation with specific repositories' APIs (Dryad, TDR, Zenodo). 'DataCite + OpenAlex APIs' represents additional Figshare datasets that are indexed in DataCite, labeled as 'dataset,' and that do not contain any affiliation metadata. The initial and post-cleaning counts are the same because the initial query is intentionally made to be affiliation-agnostic (searching for all datasets listed for a scholarly publisher), whereas all other search methods specify an affiliation and presume that retrieved records are legitimately affiliated. More than 230,000 mediated datasets of uncertain affiliation are retrieved in the initial query, most of which have no basis for inferring a connection to UT Austin. 'NCBI' represents BioProject records retrieved from NCBI. 'Crossref API' represents a broad query for objects labeled as 'dataset' with some combination of the terms

"university+of+texas+austin." Initial count represents the entire corpus of records retrieved initially, and the post-cleaning count represents the final corpus after cleaning and deduplication. For 'Specific repositories' APIs' and 'DataCite + OpenAlex APIs,' the post-cleaning count only includes records that were retained after combining with the general 'DataCite API' output and removal of duplicates (retaining those retrieved from the general API query); in other words, the post-cleaning counts reflect the records that could only be identified through these processes. Data as of July 1, 2025.

| Source | Initial count | Post-cleaning count |
|---|---|---|
| DataCite API | 76,618 | 2,782 |
| Specific repositories' APIs | 3,905 | 106 |
| DataCite + OpenAlex APIs | 134 | 134 |
| NCBI | 972 | 929 |
| Crossref API | 663,638 | 86 |
| TOTAL | 745,267 | 4,037 |

**Table 5. List of all repositories with 30 or more UT Austin-affiliated datasets.** The numerical threshold is applied for the post-cleaning counts. The initial counts reflect only records retrieved in the initial affiliation-based DataCite query, whereas the post-cleaning count includes records from all sources of data. For Figshare, the numbers of deposits recovered from the affiliation-based query and the targeted Figshare workflow are separated. The initial Figshare (primary) count includes mediated Figshare deposits that list a *publisher* known to be a Figshare partner (e.g., Taylor & Francis). The initial Figshare (secondary) count, the initial and post-cleaning counts are the same because the initial query is intentionally made to be affiliation-agnostic (searching for all datasets listed for a scholarly publisher), whereas all other search methods specify an affiliation and presume that retrieved records are legitimately affiliated. More than 230,000 mediated datasets of uncertain affiliation are retrieved in the initial query. Figshare and Figshare+ are combined here for both counts. For Dryad, the initial count is lower than the post-cleaning count because one dataset was not retrieved from DataCite's API due to UT Austin metadata only being in the Dryad API (not crosswalked). 'Env. Mol. Sci. Lab.' is the Environmental Molecular Science Laboratory at the Pacific National Northwest Laboratory. Data as of July 1, 2025.

| Repository | Repository type | Initial count | Post-cleaning count |
|---|---|---|---|
| Texas Data Repository | Institutional | 71,916 | 1,328 |
| NCBI | Domain | 972 | 929 |
| Dryad | Generalist | 407 | 408 |
| Zenodo | Generalist | 713 | 400 |
| Harvard Dataverse | Generalist/Institutional | 1,163 | 289 |
| Figshare (secondary) | Generalist | 134 | 134 |
| ICPSR | Domain | 242 | 84 |
| ENCODE | Domain | 503,426 | 82 |
| MassIVE | Domain | 74 | 74 |
| DesignSafe | Domain | 172 | 49 |
| Env. Mol. Sci. Lab | Domain | 1,469 | 42 |
| Figshare (primary) | Generalist | 193 | 20 |
| All other repositories | Mixed | 160,504 | 206 |
| TOTAL | | 741,385 | 4,037 |

## 3.1. General results (DataCite)

Because of the 'exact string match' nature of the DataCite API, all records returned by this workflow's affiliation-based query are guaranteed to be properly affiliated, rather than 'false positives,' which often occur for the Crossref API (see further below) and which can occur when using certain software packages like *rdatacite*, as previously mentioned. The disparity between the initial and post-cleaning counts recovered from the broad affiliation-based DataCite API query is largely attributable to file-level DOI assignments in Dataverse-based installations. There are nearly 72,000 affiliated DOIs recovered for TDR, despite fewer than 1,400 datasets proper; file-level DOIs account for more than 90% of all of the UT Austin DOIs retrieved from this query. The aforementioned over-granularization of deposits in the Environmental Molecular Science Laboratory repository also contributed to a marked disparity. Consolidating EMSL deposits dropped the unique count from over 1,450 DOIs to 42 discrete sets. The remaining gaps are attributable to the cleaning process (e.g., consolidation of multiple DOIs for the same deposit) and are discussed further below.

     A total of 65 unique repositories with datasets were identified in this corpus of DataCite records. There are nearly 80 listed in the original unprocessed output, but some were consolidated because they represent permutations of a single repository's name listing in DataCite (e.g., 'Dryad Digital Repository' vs. 'Dryad'), are intentionally labeled to not list the repository (e.g., 'Taylor & Francis' deposits on Figshare), or a repository allows users to edit the *publisher* field in a way that obscures the hosting repository (e.g., Zenodo allows freeform entry in this field); the latter field is locked to a single consistent value in the majority of repositories. All repositories with 30 or more distinct deposits are listed in Table 5. Many of these most commonly identified (not necessarily the most commonly used per se) repositories will likely be familiar to readers. TDR, our institutional data repository, has the most identified deposits, followed by NCBI, a well-known specialist repository for bioinformatics data, and a number of generalists: Dryad, Zenodo, Harvard Dataverse, and Figshare. As with Johnston et al. (2024), certain generalist repositories are conspicuously absent (e.g., OSF). Rounding out the list are a number of specialist repositories. DesignSafe (Rathje et al., 2017, 2020, 2025) is an NSF-funded repository for natural hazards and is maintained at the Texas Advanced Computing Center (TACC), a subsidiary of UT Austin. In previous iterations of this workflow (e.g., Gee & Shensky, 2025), the Digital Porous Media Portal (then the 'Digital Rocks Portal'; Prodanović et al., 2015, 2023), another TACC-managed repository, also returned several dozen results, but in recent runs, it was noted that during a platform migration, there appears to have been an issue with metadata transfer such that some previously affiliated deposits now lack affiliation metadata in DataCite; these were not manually re-added, so this repository does not appear among the highest usage repositories here. ICPSR (Inter-university Consortium for Political and Social Research; maintained by the University of Michigan) is a well-known, relatively large (>22,000 records) specialist repository for social science research. MassIVE is a mass spectrometry repository maintained by the University of California, San Diego. Finally, the Environmental Molecular Sciences Laboratory is a Department of Energy-funded facility focusing on biological and environmental sciences.

## 3.2. General results (Crossref)

For the affiliation-based query in the Crossref API, the disparity between the initial and post-cleaning counts is attributable to the nature of the API, which does not use an exact string match or a Boolean search. Most of the more than 650,000 DOIs that are retrieved are false positives that are not affiliated with UT Austin and instead are affiliated with another UT system campus, with a smaller proportion related to the use of 'Austin' in the affiliation, which includes organizations not based in Austin, TX (e.g., Stephen F. Austin University; Department of Medicine, Austin Health, University of Melbourne). I manually examined results for which none of the queried permutations of UT Austin were identified to see if there was another formulation that should be added to the config file but did not identify any. The Crossref output was then restricted to only those with a clear UT Austin affiliation, which represented less than 0.1% of the initially retrieved records ($n$ = 553; Table 6).

**Table 6. List of all repositories with 50 or more purportedly UT Austin-affiliated datasets retrieved through Crossref.** The numerical threshold is applied for the initial counts. A total of 44 platforms were included in the retrieval (some are name duplicates); Entries recorded as 'H1 Connects' and 'Faculty Opinions Ltd' are combined. 'BCO-DMO' refers to the Biological and Chemical Oceanography Data Management Office. Entries listed as 'Wiley' are thought to be Authorea preprints, but this is based on random spot-checking rather than comprehensive examination. Data as of July 1, 2025.

| Repository | Repository type | Initial count | Post-cleaning count |
|---|---|---|---|
| ENCODE Data Coordination Center | Domain | 503,426 | 82 |
| H1 Connect | Not data repository | 155,419 | 0 |
| Wiley | Not data repository | 1,866 | 0 |
| BCO-DMO | Specialist | 902 | 2 |
| EMBL-EBI | Specialist | 793 | 0 |
| USDA Forest Service | Specialist | 322 | 1 |
| Jyvaskyla University Library | Institutional repository | 201 | 0 |
| American College of Radiology | Not data repository | 171 | 0 |
| CABI Publishing | Not data repository | 74 | 0 |
| Boise State University | Institutional repository | 82 | 1 |
| The Hong Kong University of Science and Technology Library | Institutional repository | 57 | 0 |
| All other repositories | Mixed | 250 | 0 |
| **TOTAL** | | **663,638** | **86** |

    Just nine platforms are represented in this pruned output, with H1 Connect (formerly Faculty Opinions LTD) representing the vast majority ($n$ = 458, ~83%), similar to Johnston et al. (2024); these are labeled as 'dataset' even though 'peer review' has been a supported resource type in the metadata schema since 2018 (Lin, 2018). Also similar to Johnston et al. (2024), the ENCODE Data Coordination Center was listed for a decent proportion of deposits ($n$ = 82; ~15%). ENCODE is similar to NCBI (funded by the National Human Genome Research Institute), just for human genomics exclusively. For UT Austin, all of these datasets were deposited by the same researcher over a series of years, and as with Johnston et al. (2024), I

was unable to find any metadata in the Crossref schema that would facilitate deduplication or consolidation in a similar fashion to that applied for other repositories. It might be possible to retrieve additional metadata from ENCODE's REST API, but exploration of this resource is a low priority since the total count for this repository was not disproportionate compared to other repositories (unlike several thousand records for the University of Michigan in the RADS study; Johnston et al., 2024).

The remaining seven platforms had fewer than 10 deposits each. Those associated with the American College of Radiology ($n$ = 3) are case scripts, which appear to be a series of training questions. Those associated with Wiley ($n$ = 4), which are actually hosted on Authorea and which used to be labeled as such until recently, are preprints. Those associated with the Biological and Chemical Oceanography Data Management Office (BCO-DMO; $n$ = 2) are legitimate datasets. The single deposit from the Society of Exploration Geophysicists is a podcast. The single deposit from Boise State University is a legitimate dataset. The single deposit from the NumFOCUS — Insight Software Consortium (ITK) is a poster deposit on Zenodo describing a Python library. The single deposit from the USDA Forest Service is a legitimate dataset in their Research Data Archive. Those that are not definitively datasets were removed.

## 3.3. Results from DataCite-specific-repository cross-validation

The total volume of datasets recovered only through cross-validation between DataCite and a repository's own API was relatively small (Table 7), but the process revealed some nuances that may prove important for potential reusers. For example, while I was mostly interested in "gaining" datasets (i.e. datasets not found by the DataCite query but found by the repository query), there were also instances of the inverse in which a dataset was found by the DataCite query but not by the repository query. A few common cases are noted in this section.

**Table 7. Counts of affiliated datasets that were retrieved only from repository-specific APIs as part of the cross-validation step.** Counts represent post-cleaning counts (i.e. removal of any dataset that was already retrieved from the DataCite API). Data as of July 1, 2025.

| Repository | Count |
|---|---|
| Dryad | 1 |
| Texas Data Repository | 19 |
| Zenodo | 86 |
| **TOTAL** | **106** |

The first case is a lack of updated or crosswalked metadata for very old deposits. For example, some very dated deposits in Dryad (e.g., McTavish et al., 2015) and Zenodo (e.g., Dixon, 2014) have affiliation metadata recorded on the dataset landing page and in the repository API, but, as of this preprint, this metadata has not been crosswalked to the DataCite entry (sometimes the affiliation is even ROR-standardized). Identified occurrences of these deficiencies were rare, but they are likely also harder to find in any platform since affiliation metadata presence and quality were likely even poorer a decade ago.

The second case is when a DOI appears to have been published at some point but was then taken down, without any apparent indication in the API or landing page for the reason.

Rather than redirecting to the doi.org page indicating the DOI was not found, these DOIs redirect to a page from the specific repository indicating that the dataset was not found. This was observed for Dryad (e.g., Nandakumar et al., 2020b; Larter & Ryan, 2023) and TDR (Hodges, 2019b). For Dryad, it is presumed that these datasets were de-accessioned (temporarily or permanently) for an undeclared reason; this repository does not have a tombstone process for such datasets. Curiously, there appears to be a dataset with nearly identical metadata and a public DOI (Nandakumar et al., 2020a) for one of the non-functional Dryad DOIs (Nandakumar et al., 2020b). For TDR, the explanation is less clear (Dataverse installations do have a public-facing tombstone mechanism), but the unavailable dataset also appears to have a duplicate, publicly available dataset with nearly identical metadata in TDR (Hodges, 2019a). The final example is a TDR dataset that is published and publicly accessible, but the DOI has not been activated (Bernard, 2017). These instances have been reported to their respective repository staff and ideally would demonstrate the utility of a workflow like this for identifying datasets in need of metadata enhancement or repair.

## 3.4. Results from cleaning and de-duplication

The cleaning process is arguably more involved than the discovery process because it requires fairly granular knowledge of individual repositories' practices of DOI minting (e.g., whether files receive DOIs, how versioning is handled, whether version-related DOIs have an obvious relationship based on their construction). For this reason, the progressive difference in counts before and after certain cleaning steps is shown here to hopefully illustrate the import of these steps, which are rooted in extensive manual examination of metadata from early outputs and exploration of repository practices. The RADS study did not explicitly perform additional deduplication for multi-DOI deposits other than for Zenodo (Johnston et al., 2024), so this demonstration is based on their dataset (Mohr & Narlock, 2024; Fig. 5).

     The significant impact of accounting for partial/complete DOI duplicates and over-granularization of some mediated deposits is demonstrated in Figure 5; for most RADS institutions, the final count is less than 15% of the original reported dataset volume. For repositories that mint a parent DOI that resolves to the most recent version and a DOI for each version, proper deduplication should result in at least a 50% reduction in the number of records but can produce a greater reduction if versioning is common. The consolidation of Figshare deposits that were over-split by the automated process further reduces the number of deposits that are considered to be truly distinct. This exploration is not meant as a criticism of Johnston et al.'s (2024) methods — it may be preferable to overestimate than to underestimate in most scenarios of this work — but it underscores the importance of critical examination of a given dataset and accounting for specific repositories' nuanced practices, especially if analyses of such data may be used for key decision-making processes. As noted in the Methods, the optional consolidation of datasets in Dataverse in a similar fashion was tested but not implemented in the iteration that is presented here.

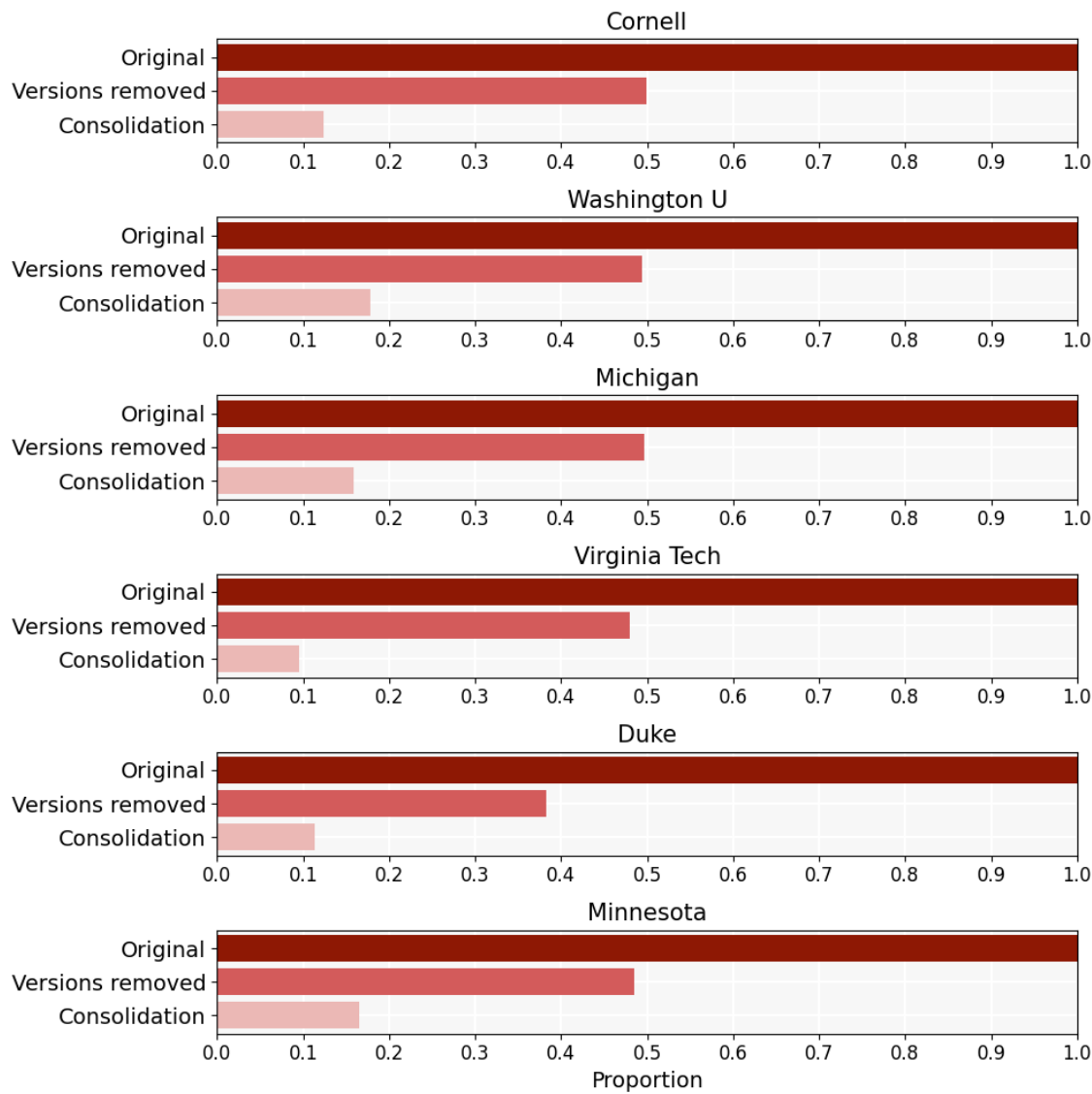

**Figure 5. Comparison of reported dataset counts for RADS institutions from Johnson et al. (2024) versus counts after removal of redundant version DOIs and consolidation of Figshare dataset deposits.** The relative proportion is scaled against the total number of Figshare datasets for each institution (when a dataset has multiple RADS affiliations, it is counted for each one). 'Versions removed' removed any DOI that ended in '.v*'; this should remove all but the 'parent' DOIs and should account for both a lineage of versions where the RADS affiliation was present in all versions and a lineage in which the affiliation was only present in more recent versions. In theory, this proportion should be no more than 50% but could be lower (e.g., Duke) if there are some datasets with more than one version. 'Consolidation' deduplicated entries that shared an *institution*, *publicationDate* (MM/DD/YYYY format), and the entire *relatedIdentifier* list or dictionary; this represents the number of unique articles represented among these Figshare deposits and returns counts equivalent to those returned by this workflow, which accounts for over-granularization of some mediated Figshare deposits (i.e. one DOI for each file associated with the same manuscript). Institutions with relatively low proportions retained after this process (e.g., Duke, Virginia Tech) have a relatively high file:article ratio (i.e. highly over-split deposits). The summary table with exact counts is included in the dataset associated with this preprint (Gee, 2025b).

A final observation made during the cleaning process was the rare duplicate publishing of the same dataset in different repositories; presently, three pairs affiliated with UT Austin are identified: TDR + Zenodo (Samineni, 2022; Samineni & Kumar, 2022); Dryad + ScienceDB (Wang et al., 2023, 2023b); and Harvard Dataverse + Zenodo (Ishikawa et al., 2025a, 2025b). The reasons for this duplication are unclear, but with their identification, we would be able to theoretically contact the researchers to inquire about this practice. Also of note is the presence of many paired Dryad-Zenodo datasets that share the same title and most other metadata fields; these result from the partnership between these repositories in which a researcher submitting to Dryad has the option to create a linked Zenodo deposit for supplemental information or software (Lowenberg, 2021). Linked Zenodo deposits will contain separate files and be licensed differently, but other metadata attributes like title, authorship, and affiliations are shared with the Dryad deposit, and these deposits are not treated as partial duplicates since they represent an appropriate splitting of materials for one project.

The Crossref cleaning process was relatively straightforward for UT Austin because very few deposits were truly affiliated with the institution. The primary cleaning step is in searching for a permutation of UT Austin and then manually checking unmatched deposits to see if there is another permutation not currently in the queried list. None of the retained affiliated deposits or repositories exhibit the types of partial duplication (e.g., DOIs for each version in Figshare or Zenodo) that was observed for deposits minted through DataCite, so no additional deduplication or consolidation steps were taken. In the future, more attention may be given to investigating whether ENCODE deposits can be partially consolidated. If a process can be effectively designed to capture the millions of mediated Figshare deposits with DOIs minted through Crossref, a process to consolidate related but separate DOIs (e.g., fifteen supplemental tables for one PLOS paper) will be essential and warrant a comparative analysis.

## 3.5. Results from targeted Figshare workflows

There are nearly a dozen known (current and former) publisher partners who mint their mediated Figshare DOIs through DataCite. Some of these have only a few dozen to hundred total Figshare deposits or are narrow in publishing scope; thus, some publishers do not appear to have any mediated Figshare deposits associated with UT Austin. Taylor & Francis records the largest number of articles with at least one associated UT Austin-affiliated Figshare deposit. It is also likely that publishers are generating mediated deposits that are labeled as something other than 'dataset,' especially in journals that cover disciplines that are not historically regarded as data-producing (e.g., humanities and fine arts). Although the number of additional datasets gained through this process is not proportionately large (<200 datasets, ~4.5% of the currently recorded total across all repositories), it is a six-fold increase in the number of distinct Figshare deposits after accounting for the over-granularization of mediated Figshare deposits that were recovered through the initial affiliation-based DataCite query (20 vs. 134, or a 670% increase in the number of affiliated Figshare deposits; Fig. 6; Table 5).

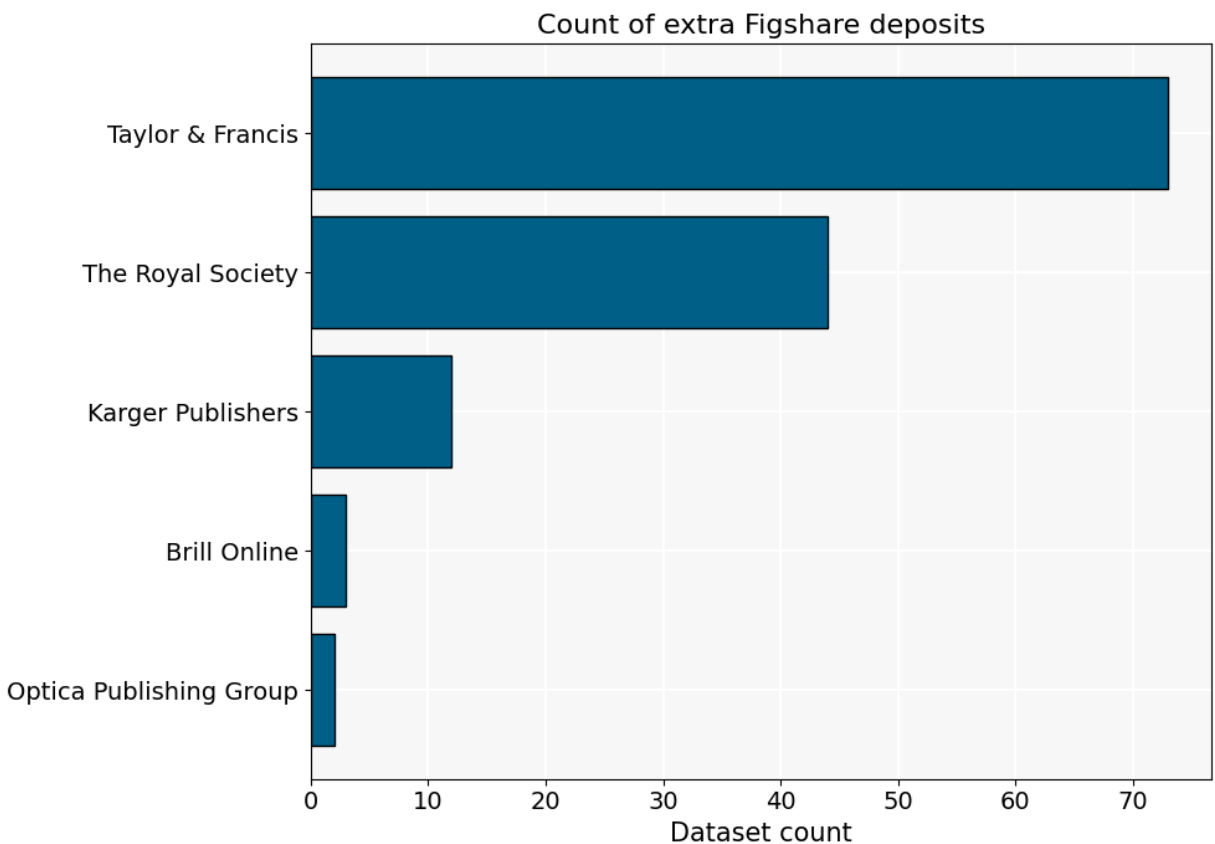

**Figure 6. Comparison of the number of additional Figshare DOIs without affiliation metadata that were discovered through connections with affiliated articles.** Note that not all publisher partners are represented here because some mint DOIs through Crossref, and their deposits cannot be discovered in the same way, and not all partners who mint DOIs through DataCite are large-volume publishers (i.e. there do not appear to be any UT Austin-authored articles that deposited supplemental information in those publishers' titles). How these mediated deposits are classified is likely also an explanator. Counts represent post-deduplication counts (*n* = 134). Data as of July 1, 2025.

     Given the sheer volume of mediated Figshare deposits created by publisher partners who use Crossref for the DOIs (on the order of millions), an efficient solution to identify all mediated DOIs remains elusive. Only preliminary results from the PLOS web-scraping workflow are described here to give an idea of the scale of these deposits. OpenAlex returned 636 articles that are authored or coauthored by UT Austin researchers and that have at least one Supporting Information file. Between those articles, there were 3,879 files published as Supporting Information files, or an average of more than six SI DOIs per article. The highest number of DOIs associated with one article was 38. This volume underscores both the difficulty with using a scripted approach to retrieve them all from Crossref and the inflated nature of dataset 'counts' from this source. These are distributed among a limited set of options that PLOS provides for labeling SI files (e.g., Appendix, Audio, Checklist, Code, Data, Dataset, Figure (and Fig.), File, Movie, Protocol, Table, Text, Video), which do not correspond to their indexing in the Crossref metadata ('component' is the typical label based on random examination of DOIs), so even with the ability to obtain this information for PLOS

articles, there is still no reliable method for discerning which ones are datasets, mixed deposits with data, or definitively not data, further complicating attempts to utilize this source of datasets. For this reason, these records were not integrated into the main corpus that is analyzed here.

## 3.6. Software publishing

Making a DataCite query for objects labeled as 'software' with any one of the many UT Austin permutations returned just six unique repositories, with nearly all of the approximately 400 unique (post-cleaning) deposits being published in Zenodo (Table 8). As noted in the Methods, based on the number of purported UT Austin-affiliated GitHub repositories (Gee & Shensky, 2025), I suspect that only a minute fraction of research software is being published in the scholarly sense (with a DOI or equivalent PID) and labeled as a 'software' object (Crossref does not support this resource type). As will be shown in the next section, at least some amount of software is being published either with data or as 'datasets.' The present results for objects classified as 'software' are relatively uninformative beyond telling us that when researchers publish software and have a correctly labeled deposit, they overwhelmingly do so through Zenodo.

**Table 8. Summary of affiliated software publications from the general affiliation-based DataCite Query.** *The original entry that was reclassified as being from the Department of Energy's CODE repository originally had the publisher listed as 'University of Texas at Austin,' the affiliation of the second author; manual inspection revealed the proper repository listing for this deposit. Data as of July 1, 2025.

| Repository | Initial count | Post-cleaning count |
|---|---|---|
| Zenodo | 1,416 | 366 |
| Code Ocean | 27 | 20 |
| brainlife.io | 3 | 3 |
| Sandia National Laboratory | 2 | 2 |
| Department of Energy CODE | 0* | 1 |
| Pacific Northwest National Laboratory | 1 | 1 |
| **TOTAL** | **1,450** | **393** |

## 3.7. Metadata assessments

One of the potential inferences that can be derived from any dataset of institutionally affiliated datasets is which platforms are preferred by researchers. However, given the frequency of external collaborations, the specific role of affiliated researchers in any aspect of data publishing, from which repository to use to the quality of metadata, is not always apparent. In the absence of a system like CRediT or MeRIT (which remain far from universal adoption in journals) and the potential for individuals to have "contributed" to, rather than "authored" an output (e.g., Holcombe, 2019), this workflow uses the presence of an affiliated researcher in the first and/or last author positions as a crude proxy for characterizing whether an affiliated researcher likely played a role in repository selection (Fig. 7).

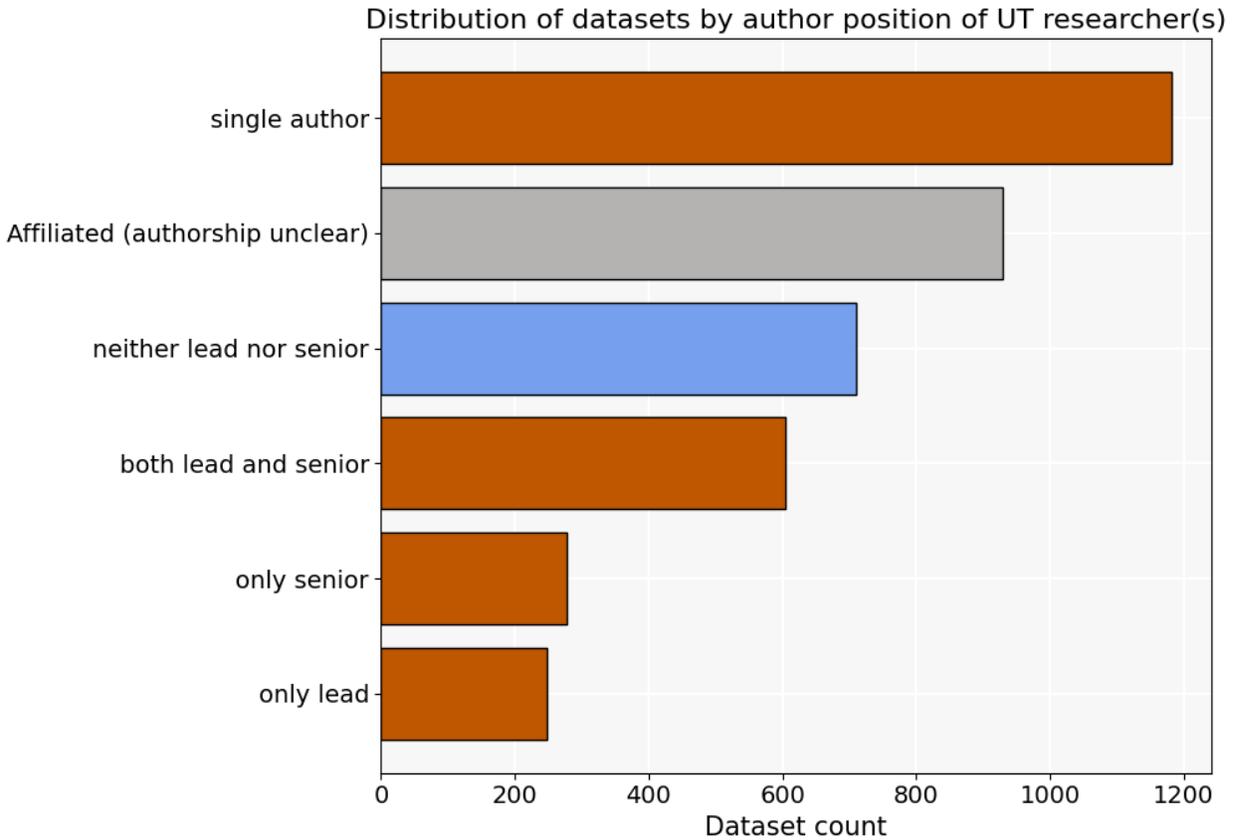

**Figure 7. Comparison of authorship position of UT Austin authors on affiliated datasets.** Orange demarcates any category in which a UT Austin researcher is lead and/or senior author; blue demarcates when a UT Austin researcher is neither lead nor senior; and gray (only NCBI) results from the different metadata schema used for their records, in which specific authorship is less clear. The data depicted represent all datasets across all sources (*n* = 4,037). Data as of July 1, 2025.

Examining specific repositories through this lens reveals some interesting patterns (Fig. 8); for example, UT-affiliated Dryad deposits are less likely to have a UT Austin author in the lead and/or senior position than TDR or Harvard Dataverse. This is somewhat intuitive as Dryad charges data publishing charges if the cost is not covered by a partner journal or member institution (which UT Austin is not), so UT Austin researchers are probably unlikely to choose Dryad unless this fee is covered as part of a journal sponsorship or institutional membership of a coauthor. However, Zenodo deposits also show a relatively show frequency of a UT Austin author in the lead and/or senior positions, despite being free to deposit, so cost is not the sole explanator; other considerations such as collaborators' preferences may play a role in the relatively low frequency. Interestingly, the overwhelming majority of UT Austin-affiliated deposits in Harvard Dataverse (free for all depositors) do feature a UT Austin researcher in the lead and/or senior position; the proportion is quite similar to TDR.

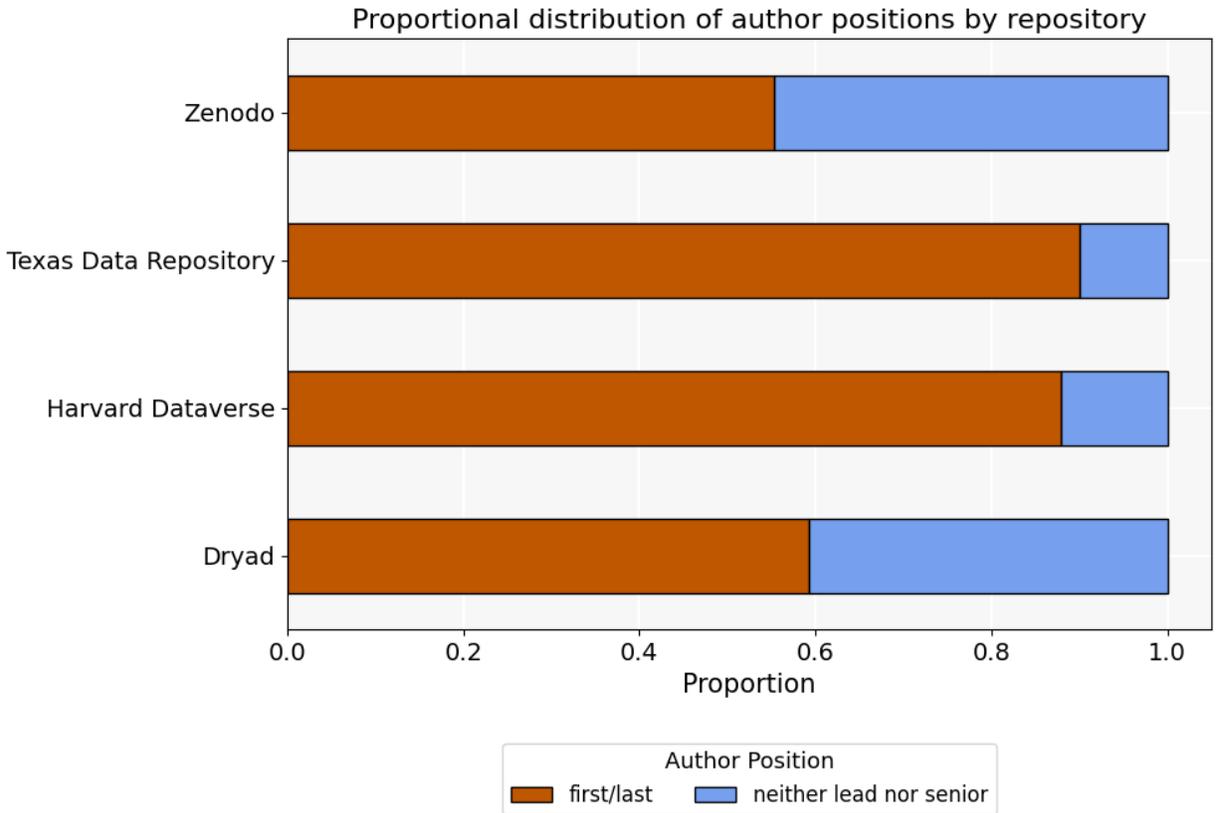

**Figure 8. Comparison of the frequency of datasets with a UT Austin researcher listed in the lead or senior author positions among select repositories.** The selected repositories are based on those with the highest number of UT Austin-affiliated datasets (excluding NCBI, which does not record authorship in the same way). Orange includes any category depicted in Figure 10 in which a UT Austin researcher is lead and/or senior author; blue demarcates when a UT Austin researcher is neither lead nor senior. The data depicted represent all datasets across all sources (*n* = 4,037). Data as of July 1, 2025.

Other attributes beyond authorship can also be explored based on a combination of existing metadata fields and metadata that can be derived from existing metadata. Two other examples are briefly discussed here. Firstly, the frequency at which 'datasets' contain software, or consist only of software, is an important metric as part of a coarse-level assessment of whether objects are properly classified in metadata schemas. For UT Austin, 261 datasets from the initial DataCite query retrieval include one or more software formats in addition to a non-software format (Fig. 9), with 22 datasets comprising exclusively software. While this might seem like a relatively small proportion, it bears noting that many repositories do not crosswalk file format metadata (*n* = 1,232 datasets without this information), so the 283 datasets with at least one software format actually represent about 17% of the sampled records that do crosswalk this information, or about one in six datasets. Presumably, some compressed archives also contain software files but require targeted exploration to verify. This observation speaks to both the frequency of mixed-type deposits and the challenges associated with querying records based on object type in the DataCite schema.

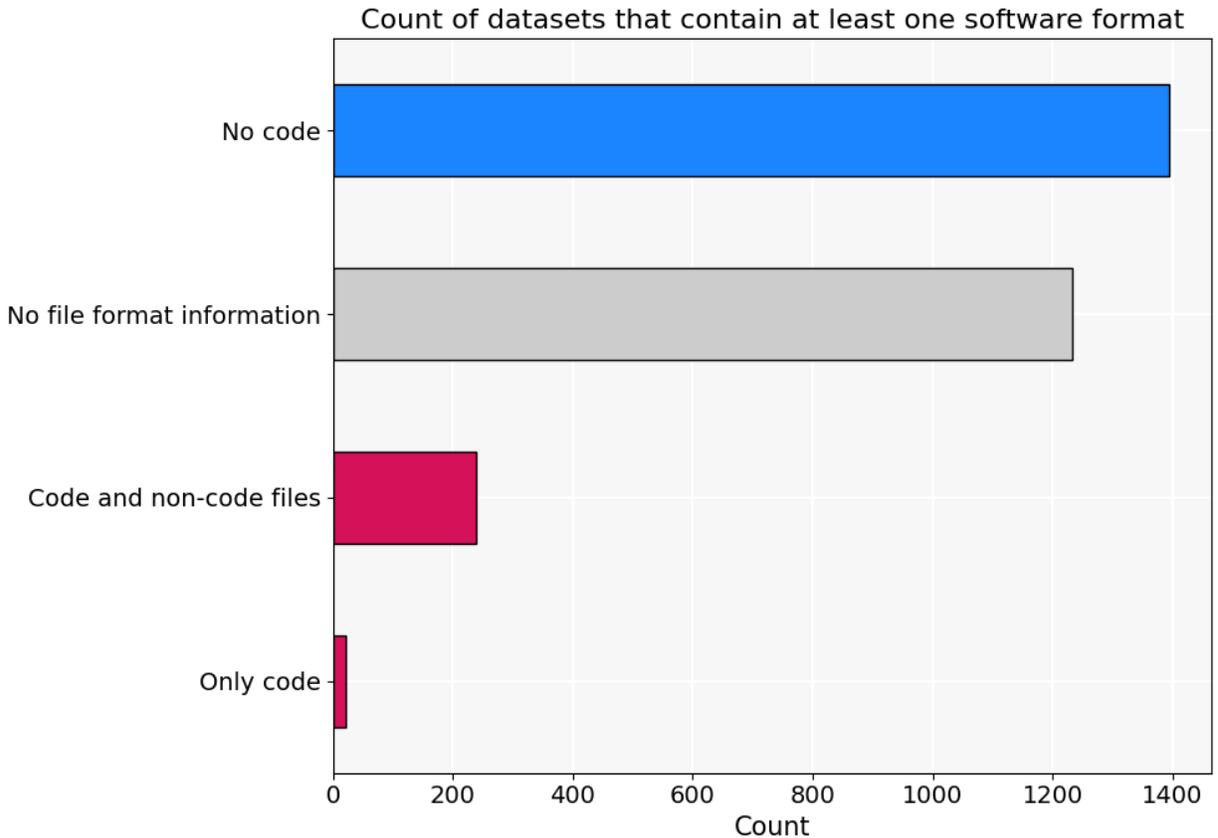

**Figure 9. Count of affiliated datasets based on whether one or more software formats was detected in a dataset's metadata.** The data depicted represent all datasets (excludes deposits labeled as 'software') retrieved in the general affiliation-based DataCite query and the cross-validation with specific repositories' APIs (*n* = 2,888); mediated Figshare deposits with DataCite DOIs are not included. Red bars represent datasets with at least one software format; blue datasets do not contain a (readily identifiable) software format, although software could be present in more generic file formats (e.g., .txt files). Similarly, just because a non-code format is present does not mean it is 'data.' As can be seen in the gray bar, many datasets do not contain any information on file formats (this is sometimes independent of whether information on file or total deposit size is provided). Data as of July 1, 2025.

     Licensing is a common challenge for researchers and is intertwined with mixed-format deposits since the ideal licensing for data and software often differs. This is evidenced through a small number of 'dataset' objects with software licensing, such as Apache, BSD, and MIT (Fig. 10). Although initial examination of the distribution of licenses among datasets appears to indicate that the CC0 license waiver / public domain designation is by far the most common (a positive sign since this is often considered the most appropriate treatment for data *sensu stricto* in the United States), it bears noting that some repositories only use CC0 licensing (e.g., Dryad) or have CC0 set to the default, requiring researchers to manually select or enter non-CC0 terms if they so desire (e.g., Dataverse installations). The number of Dryad and TDR deposits that presumably use CC0 because researchers were required to, were not aware of their ability to change the license, or simply did not care about the license selected, is nearly 1,700; if these repositories are omitted, CC0 drops to the third most

common license treatment behind the aggregate category of unclear rights and CC BY. This shift suggests that researchers are not necessarily actively choosing the optimal treatment for data of their own volition. Similarly, CC licenses may be slightly inflated by the automated Figshare workflow (not included in Figure 10), which often assigns a CC BY or CC BY-NC license based on personal observations.

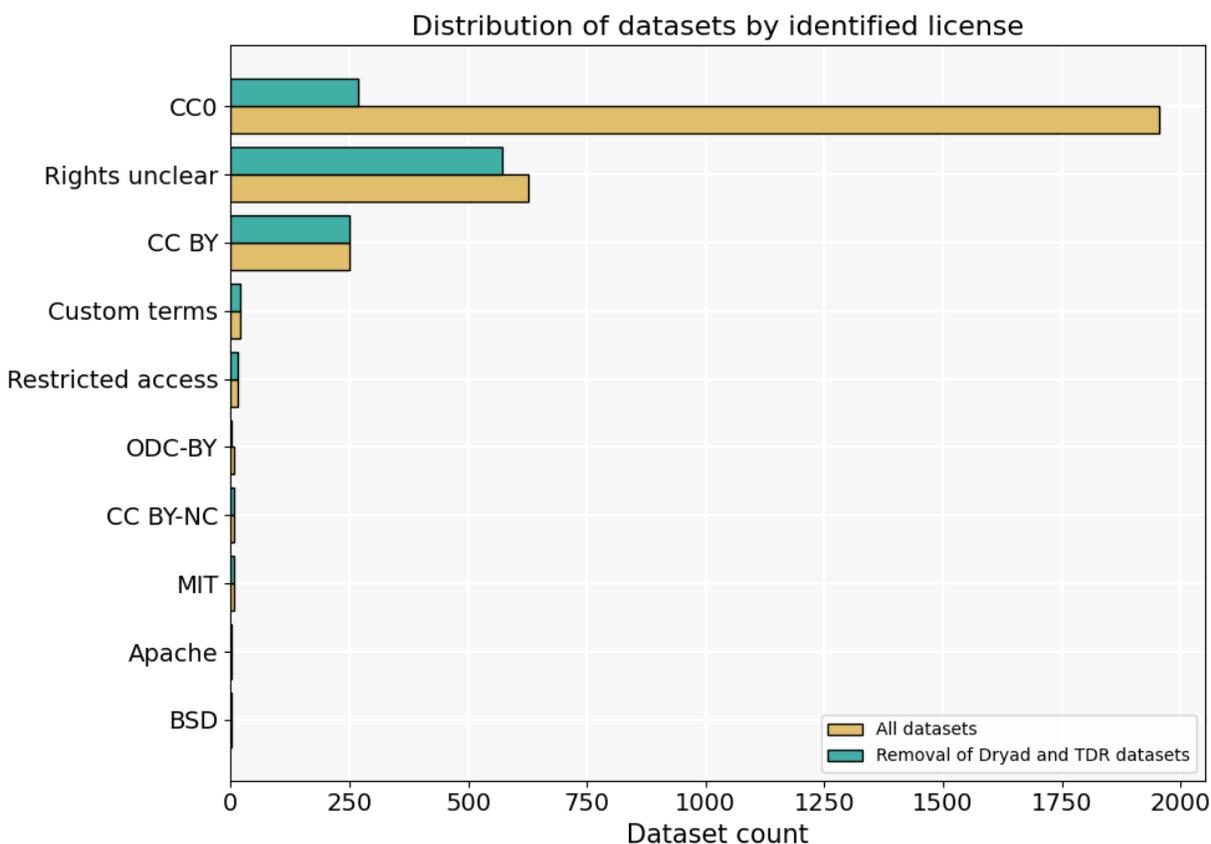

**Figure 10. Comparison of licensing treatment for objects labeled as 'datasets.'** Data are based on the *rightsIdentifier* field values and are standardized. The data depicted represent all datasets retrieved in the general affiliation-based DataCite query and the cross-validation with specific repositories' APIs; mediated Figshare deposits with DataCite DOIs are not included (*n* = 2,888). Deposits classified as 'rights unclear' could have this information recorded (e.g., dataset landing page), but it was not crosswalked into the DataCite API. Data as of July 1, 2025.

# 4. Discussion

## 4.1. UT Austin snapshot

Although this paper focuses more on describing the development of the workflow and general challenges to institutional discovery, the primary objective of this work at UT Austin is to use the data to identify trends for our campus specifically and then to use that to inform the services and support that the UT Libraries provides to researchers on our campus. This section provides some examples based on analysis of the present corpus of data.

#### 4.1.1. Comparison among repositories

Predictably, the most utilized repositories for discovered datasets are generalists, both our institutional data repository (TDR), and large, well-known generalists like Dryad, Harvard Dataverse, and Zenodo. The relatively sustained usage of Harvard Dataverse is particularly interesting to us since TDR is built on the same software and has nearly all of the same features, so there is not a clear "advantage" of using the former compared to the latter in the same way that a researcher might prefer a non-Dataverse generalist for certain features. As previously shown in Figure 8, most of the UT Austin-affiliated deposits in Harvard Dataverse have a UT Austin researcher in the lead and/or senior position, so the relatively high dataset count does not appear to be the result of collaboration with external researchers who prefer the platform.

A longitudinal comparison of the five repositories with the highest number of UT Austin-affiliated deposits also reveals some interesting differences. Three of the five repositories show some measure of an increase in annual deposits over time (Harvard Dataverse, TDR, and Zenodo), while two are flat in recent years (Dryad and NCBI). It is not possible to discern the relative influence of journal data sharing mandates; funder mandates; and a general increase in the volume of published articles from these data alone, but some hypotheses for the observations can be proposed.

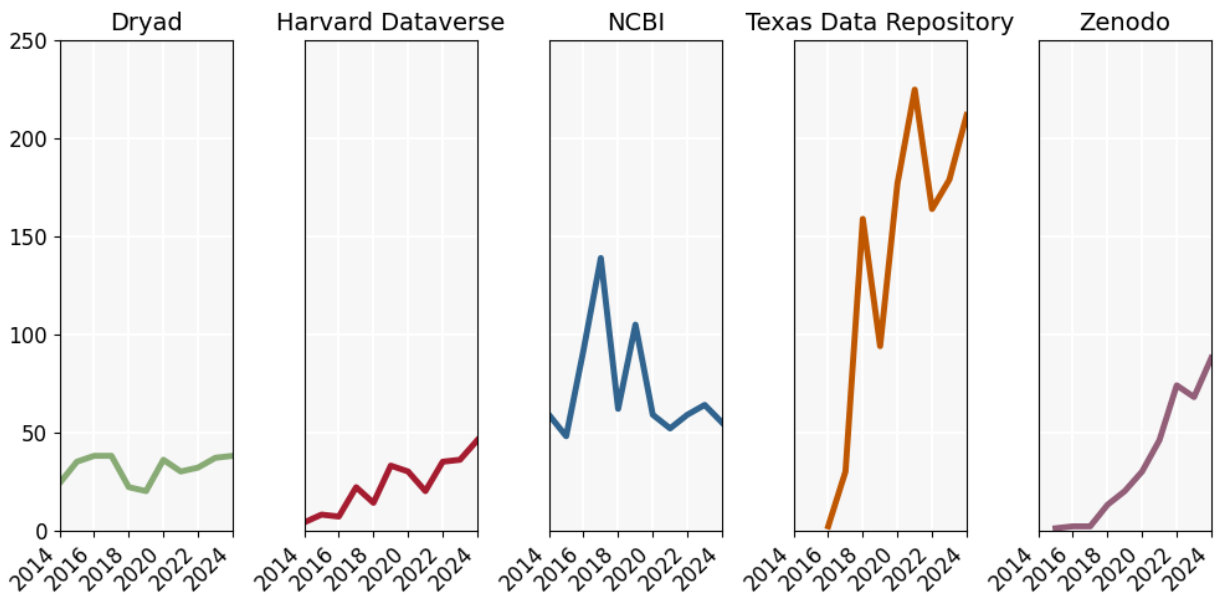

**Figure 11. Comparison of the annual volume of UT Austin-affiliated dataset publications in select repositories among select repositories (2014–2024).** The selected repositories are based on those with the highest number of UT Austin-affiliated datasets. The data depicted were drawn from the corpus of all datasets across all sources and excludes objects labeled as 'software' in the DataCite schema (mainly relevant for Zenodo numbers, which would be much higher if software deposits were included). Data as of July 1, 2025.

The three repositories with a sustained increase in UT Austin-affiliated deposits show slightly different trends (Fig. 11). Harvard Dataverse's pattern can best be described as a slow linear increase, with relatively constrained oscillation. By comparison, TDR, which only began accepting deposits in earnest in 2017, is characterized by a steeper but more erratic linear increase. More pronounced year-over-year volatility is likely linked to the relative over-granularization of sets of materials (when researchers create many separate 'datasets' for a single paper that are housed within a single 'dataverse' object), although this practice may be warranted in some situations. Although the same over-granularization is possible in Harvard Dataverse, the practice appears to be far less common for reasons unknown. Finally, Zenodo is characterized by an increase closer to an exponential curve than a linear one, going from nearly zero deposits to about 75/year (with additional, sub-equal volume for Zenodo software deposits). This rise may relate to the lack of early awareness of Zenodo as a European-based platform given the presence of other established generalists used by U.S. researchers. Recent developments that could have increased Zenodo's profile include a partnership with the U.S.-based Dryad (early 2021; Lowenberg, 2021) to mediate software materials to Zenodo as part of data submission to Dryad; increased usage of GitHub, which features an integration with Zenodo to automatically mint DOIs for GitHub repositories (the integration has existed for over a decade, but GitHub usage by researchers has significantly increased in recent years; Färber, 2020; Escamilla et al., 2022, 2023; GitHub, 2024); and involvement with GREI (early 2022). Zenodo also has one of the largest file size limits among generalist repositories (50 GB), which is significantly more than Dataverse-based repositories and OSF (≤5 GB).

The two high-volume repositories without an increase in UT Austin volume also exhibit differing trends. NCBI deposits peaked several years ago, declined, and now appears to have plateaued. Because NCBI metadata is not constructed in the DataCite schema (e.g., does not record individual authors and affiliations in the same fashion), it is more difficult to discern the full scope of the research team and institutions involved with the data. Conceivably, there are deposits where the collection and/or analysis may have been done at UT Austin, and the deposit lists UT Austin, but the overall project was not led by a UT Austin researcher (or vice versa). It would be interesting to compare the trend in NCBI deposits to trends in NIH and NSF funding for UT Austin researchers, but this is beyond the present scope of the workflow. Publication in Dryad is essentially flat over time, which reflects a similar trend across total deposits in this repository. For UT Austin, which is not an institutional member, affiliated researchers are less likely to choose Dryad unless the data publishing charge (DPC) is covered by a sponsoring journal to which the associated manuscript is submitted, which is observed in Figure 8. Therefore, the UT Austin trend for Dryad deposits likely reflects a relatively stable publishing volume in certain journals and by certain research groups, suggesting that Dryad has "cornered" a certain market share but is not expanding that share, despite its increased profile through GREI and other partnerships.

### 4.1.2. Metadata assessment

The development of metadata assessment processes is still in its early stages, but hopefully the examples provided in the discussion demonstrate the value of such components in the workflow. Inferring whether UT Austin researchers were actively involved in repository selection based on authorship position (Figs. 7–8), among other attributes, is important for more nuanced interpretations of bulk data such as comparisons of the total number of affiliated deposits published in different repositories. Uncritical analysis of the entire corpus is almost assuredly skewed by patterns of collaboration, authorship conventions, and disciplinary culture.

Separately, the prevalence of software in objects classified as 'datasets,' and the associated myriad of licenses applied to different deposits (Figs. 9–10), highlights an area known by many librarians to be a major deficiency in researchers' comprehension of best practices around both data and software management. As many research fields become more computational and universities become increasingly interested in software (some of which carries commercial potential), appropriate division of materials and application of licensing, and education around these topics, will become more critical.

## 4.2. FAIR for whom: open data, closed metadata

The majority of work on best practices around research data and software management and sharing focuses specifically on other researchers as the predominant use case. In this respect, it is not surprising that affiliation metadata is neither uniformly nor consistently recorded by repositories. Relative to metadata such as title, description, keyword, and subject area, author affiliation and name are less likely to be important search parameters for the average researcher because most researchers are not conducting institution-specific or bibliometric work. Nonetheless, similar to how there is a general consensus around the importance of preparing and sharing research data for purposes beyond additional/confirmatory research (e.g., educational potential of some data types), there should also be a push to facilitate findability by entities beyond other researchers. The emphasis on recording and standardizing funder metadata across repositories of all kinds demonstrates that it is possible for secondary use cases to exert such influence on metadata standards, albeit through potential monetary considerations (e.g., the GREI initiative provides funding for participating repositories).

In this preprint, I have made frequent mention of certain repositories that have made public commitments to, and received extensive federal funding for (via GREI), implementation of certain metadata standards but that do not adhere to these in practice. These mentions are made not out of any particular animus towards these repositories but rather out of an observation that (1) these are large, federally funded repositories; and (2) the guidance that they have contributed to as part of GREI is not in alignment with their practices. For example, per Van Gulick et al. (2024), "*repositories need to collect information about the dataset producer and their affiliation*." That repositories like Figshare and OSF only collect this information for a handful of research organizations that are institutional members hardly seems sufficient to qualify as having achieved this aim. There is a certain irony that repositories with an expressed commitment to open data have either closed (internally recorded but non-crosswalked) or non-existent metadata that should be considered the minimum baseline. Similarly, Curtin et al. (2023) stated that "*we also hope this common metadata schema* [the GREI schema] *will be useful for data repositories beyond GREI to improve interoperability across data repositories and across the NIH data landscape*," with the caveat that their white paper only "*strongly encourage*[s] *that each repository member collect the following metadata*." It is difficult to envision a scenario in which other repositories would feel a strong impetus to adopt GREI recommendations that are not even adhered to by some GREI repositories.

Many repositories, both generalist and specialist, offer institutional memberships as a part of their business model (e.g., Dryad, Figshare, OSF). Some of these repositories offer enhanced institutional discovery as a membership benefit, which operates on the implicit premise that it will not be possible to easily or systematically track institutional outputs in these repositories without membership because this information is either not recorded or not standardized. It seems unlikely that these repositories will significantly alter their business model since access to institutional tracking dashboards is one of the main selling points, especially for repositories that are free to deposit for non-affiliated researchers (contrasted with Dryad, for example, whose institutional membership waives data publishing charges for affiliated authors). The above commentary is not a criticism of different business models, especially given the present uncertainty around many external funding sources. Nonetheless, there are clear shortcomings for certain use cases and stakeholders that result from this opacity or deficiency in metadata. Particularly in the case of Figshare's integration with various scholarly publishers, there is both clear evidence that the current workflow is suboptimal and that improvements can be made. Some brief remarks to this point are made below.

The marked variation in how mediated Figshare deposits are created, ranging from which indexer is used to mint DOIs to how materials for a single manuscript are assigned DOIs and labeled in the metadata, is a clear impediment. Speaking from personal experience in different capacities, that this integration exists and how it functions is not well-documented. For example, in Taylor & Francis titles that I have published with, there is no indication that this integration exists in the journal instructions, submission portal, or any part of the production process; it only appears in the high-level publisher guidance, which is a less likely source of guidance for researchers compared to journal-specific guidance or specific repository user guidance. The integration presumably pulls metadata from the journal submission system (making it all the stranger that these deposits record no affiliation or ORCID information), but there is evidence (from my own publications) that some metadata is auto-generated. In many instances, this has resulted in quite poor metadata, such as overly expansive subject area designations that essentially capture all topics within a broad discipline or badly formatted keywords. For example, my anatomical description of a fossil amphibian, published in *Frontiers in Earth Science* (Gee et al., 2019a), resulted in a Figshare deposit (Gee et al., 2019b) that includes subject area designations such as 'Volcanology,' 'Natural Hazards,' and 'Inorganic Geochemistry', words and phrases not even mentioned in the article. Another anatomical description, published in the *Journal of Vertebrate Paleontology* (Taylor & Francis; Gee et al., 2023a), resulted in a Figshare deposit (Gee et al., 2023b) with poorly constructed or malformatted keywords that I did not create in any system, such as "presently known"; "example .)"; and "lydekkerina <".

As I have never logged into Figshare, and none of my auto-created deposits record even basic author metadata such as my ORCID or affiliation at the time, it seems unlikely that I would be able to claim, edit, or version any of these datasets to improve their quality (although I have noticed at least one instance in which either the journal or Figshare versioned the dataset without my knowledge let alone any request from me; compare Gee et al., 2019d and Gee et al., 2020). In my capacity as a librarian, identifying the many metadata nuances associated with the mediated Figshare process was entirely the result of extensive exploratory research, not some transparent public resource or standardized practice.

However, there are also examples of how the Figshare integration can be deployed optimally, or at least more optimally than at present. For example, The Royal Society of London is a publisher partner who mints Figshare DOIs through DataCite. The production teams for their portfolio add a full bibliographic reference and an in-text parenthetical citation for any mention of SI that is hosted as mediated Figshare content (e.g., So et al., 2024b for So et al., 2024a). This achieves at least the aim of data citation, which practically never occurs for mediated Figshare deposits (conflicting with another GREI output; Puebla et al., 2024) because most authors are not even aware that their supplemental information has become discretely citable. This demonstrates how a best practice can be routinely integrated into the workflow. In a different example, I previously noted how nearly all DataCite deposits that record affiliation metadata, regardless of whether those affiliations are institutional members, and that list Figshare as the *publisher* are associated with Springer Nature articles. I cannot assess whether this is an exclusive feature related to their organizational proximity under the Holtzbrinck Publishing Group, but the existence of these deposits for UT Austin (a non-institutional member of Figshare) demonstrates that it is possible to record and crosswalk affiliation metadata for these mediated deposits. Given that the Figshare integration has generated millions of DOIs to date, there should be an imperative to optimize this workflow.

Data repositories strive to meet the FAIR principles, and in theory, should strive to maximize FAIRness of deposits (e.g., high-quality metadata, open formats, proper licensing), rather than meeting the bare minimum (e.g., minting a DOI, the act of which carries certain requirements). The 'Findability' part of the FAIR principles requires metadata and data to be findable without *a priori* knowledge of their existence, which is one of the primary reasons why supplemental information is a problematic medium for supporting research outputs. Similar to how reusability should be thought of as possible beyond the strict confines of a particular domain or even the general research ecosystem, Findability must also be thought of beyond the confines of researchers trying to find data. If datasets are not findable at an institutional scale, I would argue that they are not really FAIR. Although significant attention is directed towards improving author behaviors around data publishing, there is a clear need for repositories to also critically assess their own practices through a lens of FAIRness as a gradient.

## 4.3. Challenges of systematically identifying specialist repositories

Both this study and the RADS study (Johnston et al., 2024) return results where a handful of repositories, mainly generalists and institutional data repositories, account for the overwhelming majority of identified affiliated deposits. Although some specialist (domain-specific) repositories are well-represented in the dataset, higher-volume ones tend to be related to specific nuances of a university (e.g., DesignSafe is more likely to be used by UT Austin researchers as a UT Austin-managed repository). It seems likely that the majority of specialist repositories are not being captured in previous and current automated processes based on the number of repositories that are identified versus the approximate number that exist; most of the thousands of registered data repositories are specialist in nature.

In some (perhaps many) instances, specialist repositories do not record affiliation metadata, do not crosswalk it, or do not use a PID system that supports centralized discovery across repositories of all kinds. This can range from high-volume, high-profile repositories such as those under NCBI, which also follows a more traditional physical repository system for issuing PIDs in that it issues accession and project numbers but does not mint DOIs, handles, or an equivalent digital PID; to small-volume, highly specialized repositories that might use an alternative PID like an ARK. Lower quality general metadata in specialist repositories might stem from a relative lack of emphasis on broad findability since many of these repositories are well-known as central locations for broad swaths of discipline-specific data such that potential reusers are likely to go directly to a given platform rather than conducting a broad search in an indexer like Google Dataset Search. In this respect, the importance of findability without *a priori* knowledge is perhaps diminished because potential reusers are starting with a constrained, targeted search, rather than searching for data in a generalist repository or a general search engine like Google Dataset Search. Incidentally, this emphasizes the need for generalists to have rigorous standards for at least general metadata since they may be considered atypical locations for specialized data.

Given the narrow scope of many specialist repositories, institutional discovery for scientometric purposes is likely even less prominent or frequent of a use case than it is for generalist repositories. However, if specialist repositories begin shifting to institutional membership models (and away from other revenue sources like federal grants), institutional discovery may become a more important use case, and there would be a stronger impetus for ensuring institutions can assess usage of a platform. The institutional dashboards employed by some generalists with institutional membership are one possibility, but an important consideration for any repository seeking to bolster institutional membership is the ability to demonstrate some level of institutional usage prior to forming a partnership. Standardized, crosswalked affiliation metadata in the DataCite schema is the most efficient method for all parties involved.

At present, digital scholars who are conducting institutional research data discovery work will have to settle with the likelihood that a great many specialist repositories are not being captured. Although this workflow is intended to be developed to target additional specialist repositories that are known to be omitted for one reason or another (e.g., does not use DOIs, highly granular affiliation metadata), development of custom secondary workflows for many repositories is simply untenable, and which specialist repositories are of greatest use and interest will vary by institution. Discovering which specialist repositories are used by a given institution is often, in part, serendipitous.

## 4.4. Metadata plasticity and re-curation

Metadata schemas and their incorporation into repositories are constantly evolving, and metadata maintenance is theoretically a perpetual process that adapts to changing infrastructure and community needs in order to ensure FAIRness. For example, a repository may begin recording affiliation metadata or may begin standardizing affiliation metadata with ROR identifiers after having done neither previously. Previous work has demonstrated that the majority of DOIs are versioned at least once in Crossref (Hendricks et al., 2020) and DataCite (Strecker, 2024), and this proportion should increase as the Make Data Count framework is adopted. Presumably, even repositories that are maximizing the current DataCite schema and PID frameworks will need to adapt their practices in the future.

However, in order to ensure FAIRness of datasets, repositories need to not only look forward in order to develop novel practices for data that are yet to come but also backward in order to maintain data and metadata that have already been published. For better or for worse, it is unrealistic to expect researchers to update datasets simply for the purpose of expanding or standardizing general metadata in schemas and with PIDs that the average researcher is minimally aware of and for which they will not necessarily see a direct benefit. Therefore, it falls on data-hosting platforms and other staff involved in curation (e.g., research data librarians with curatorial abilities in a repository) as long-term stewards of data to enhance existing metadata records through a combination of manual and programmatic approaches (re-curation; e.g., Habermann, 2023); this is not exclusive to affiliation metadata and could also include, for example, ORCIDs for authors and links to related scholarly outputs.

In the relatively short duration of this project (nine months at the time of this preprint), there have already been some detected changes to queried metadata specifically that have required modification of the workflow and caveating of analyses. Two examples of recently detected metadata changes and one example of historic metadata changes are described.

### 4.4.1. Texas Data Repository

The first case study pertains to historic suboptimal formatting of affiliation metadata in TDR; at the onset of this project, affiliations were usually crosswalked to DataCite as a string enclosed in parentheses, an unexpected permutation of the institutional name (this occurs in a few other repositories as well). This explains, in part, the initial inability to discover TDR datasets (the absence of a leading "The" being the other explanator) through the DataCite API, and because of the relative volume of TDR deposits, '(University of Texas at Austin)' became the most commonly identified permutation across all affiliated datasets once the workflow was modified to retrieve them (see archived conference presentations by Shensky & Gee [2024] and Gee [2025c]). After raising the issue with TDR's development team, this crosswalk issue was resolved just prior to the most recent conference presentation in April 2025 (Gee & Shensky, 2025). TDR deposits are now listed with 'University of Texas at Austin' (still not the official permutation), and '(University of Texas at Austin)' is now the 7th most frequently recovered permutation. As discussed in the final section of this preprint, it is hoped that the ability to implement ROR identifiers in Dataverse installations will further improve the metadata quality.

### 4.4.2. Dryad

The second case study is date metadata in Dryad; since the onset of this project, comparisons were made between repositories with respect to their annual publication volume of UT Austin-affiliated datasets through the use of the *publicationYear* field in DataCite. At that time, Dryad exhibited an essentially flat pattern over the past decade, oscillating between a narrow range of about 35 to 45 annual datasets with at least one UT Austin researcher listed (Shensky & Gee, 2024). Prior to the most recent presentation in early April 2025 (Gee & Shensky, 2025), data collection using the same workflow returned a peculiar pattern for Dryad, with a sharp drop-off to single-digit volume in 2024, but a 2025 volume that was already at the level of previous years (Fig. 12), despite being only a third of the way into the calendar year. This odd pattern was not detected for any other repositories that were captured in the workflow.

Examination of specific datasets in the DataCite API, the Dryad API, and DataCite Commons (web interface), as well as aggregate data for all Dryad datasets in both APIs, revealed that for most Dryad datasets published in 2024, the *publicationYear* field was, at some point, updated to list '2025.' At least for UT Austin-affiliated datasets, other DataCite fields such as *date.Issued* and *registered* displayed full timestamps with the correct year. This appears unrelated to the *updated* field, which almost always lists a date in 2025, probably as part of the Make Data Count process, and also appears unrelated to versioning since most other datasets published prior to 2024 have congruent timestamps in other fields and on the landing page (i.e. if *publicationYear* was affected by versioning, we would expect to see a more even distribution of discordant dates across many years). Additionally, when examining the full corpus of Dryad datasets, there remain several hundred with a *publicationYear* of 2024 (Fig. 12).

Across all Dryad datasets, nearly 13,000 have discordant publication year fields between the DataCite and Dryad APIs. There are two periods in which the annual count based on DataCite metadata is much higher than that based on Dryad metadata (2019–2020, 2025) and one period in which the inverse is true (2024), indicating that the discordance goes both ways (some datasets appear younger than they are, while others appear older than they are; Fig. 12). The *publicationYear* field is convenient for annual summaries since it does not require extracting the year from a full timestamp and is used by DataCite to create automatic summaries in its REST API and DataCite Commons outputs. In order to ensure accurate analysis, I had to pivot this workflow to use a different DataCite field from which year had to be extracted (*registered*), which is suboptimal because these should in theory be distinct timestamps that may not be equivalent at finer timescales (e.g., month, day). Until this metadata discrepancy is resolved, it is likely impacting other processes and workflows that involve Dryad records retrieved from DataCite. Discordance in date metadata is not exclusive to Dryad; there appears to be a common issue in old Figshare datasets that have been versioned in which one creation date listed for the parent DOI is the most recent date of publication (*dates.date* where *dates.dateType='Created'*) and another (*created*) is the earliest date of publication (e.g., for Prokop, 2013, with 65 versions between 2013 and 2023, both years appear in different creation date fields; dates for 'updated' are also discordant). However, some discordance over time is to be expected as repositories change their practices, whereas the current issue with Dryad metadata is systemic but without a clear connection to a changed practice.

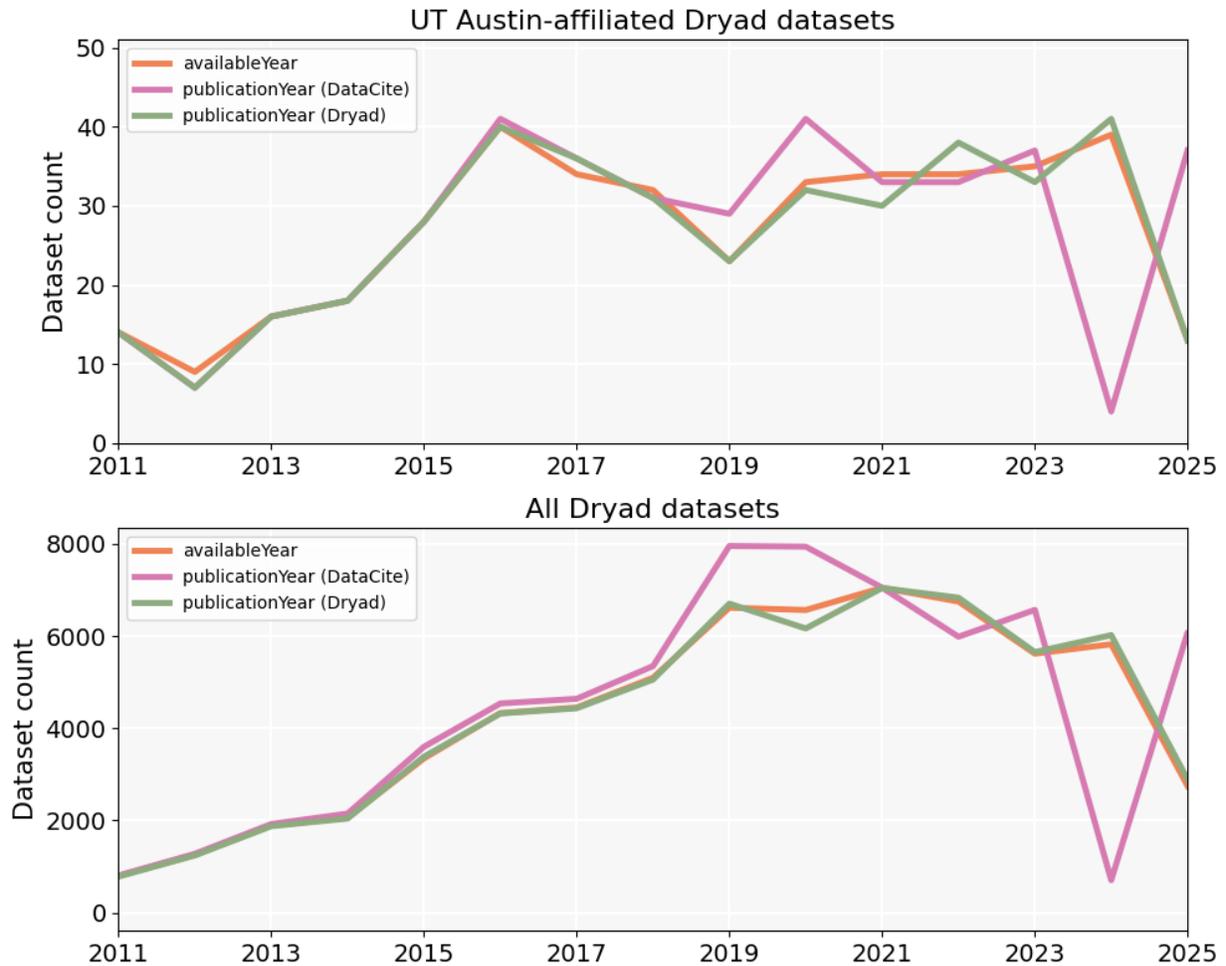

Figure 12. Comparison of the number of Dryad datasets (for UT Austin-affiliated datasets and datasets across all affiliations) with a timestamp in a given year across three timestamps. *availableYear* and *publicationYear (DataCite)* are fields derived from the DataCite API (the former is extracted from a full timestamp, the latter is taken verbatim); *publicationYear (Dryad)* is derived from a full date field in the Dryad API. These dates, particularly the two publication year fields, should be congruent (similar to the *availableYear* and *publicationYear (Dryad)* fields), but there is a marked discrepancy in 2024 and 2025, with smaller discrepancies in 2019 and 2020. This dataset omits file-level DOIs, which are retrieved from the DataCite API but not the Dryad API. Data as of July 1, 2025.

  These two case studies demonstrate the importance of both curation and re-curation of datasets to ensure their discoverability. Curation of metadata and/or data is a resource-intensive process, and many repositories, especially generalists, perform little to no curation. However, building and maintaining curation capacity and quality will reduce the need for re-curation in the future. For example, the mediated Figshare process continues to generate thousands of datasets lacking affiliation metadata, even though they are expressly linked to articles with affiliations; not addressing this now only increases the burden to do so in the future. Even in repositories that do not perform curation, it is still possible to "passively curate" through a combination of infrastructure and workflow standards in the publication process. For example, authors submitting to Dryad must select a ROR-linked affiliation from a dropdown list or must manually check a box that they did not find their institution, which can prompt a researcher to spend the time looking for their institution (Dryad does do curation and thus has a second layer of checks if an author does not select the ROR entry for an institution when one exists). As a second example, Dataverse installations can incorporate an external vocabulary that pulls the entire ORCID registry, allowing users to add co-authors with their ORCIDs already linked, rather than having to manually, separately enter names and ORCIDs; note however, this form of entry is not enforced and is not implemented in all Dataverse installations. Other examples of ways to enforce a measure of metadata quality without human curation of individual deposits can include additional required fields like a subject area (common in generalist repositories); use of controlled vocabularies (e.g., for subject area classification); minimum word counts for fields like titles, descriptions, and keywords (which could omit non-descriptive words like grammatical articles); and processes to search for, and link to, other scholarly outputs like preprints or articles, either during or after dataset publication.

However, even the best repositories will inevitably be faced with deficient metadata at some point as the infrastructure evolves, underscoring the importance of not merely maintaining but also enhancing historical records. In the context of institutional discovery, re-curation is essential when targeting records across any appreciable time period; at present, work such as this project requires knowledge of extensive caveats to the recording, crosswalking, and standardization of affiliation metadata, which, if unaccounted for, can lead to misinterpretations of the data. Mendeley Data is a good example of this: historically, it only crosswalked affiliation metadata for institutional members, even though this metadata is recorded on dataset landing pages. Based on data for UT Austin (which is not a member), it started crosswalking affiliation metadata to DataCite, regardless of institutional membership, in 2023, likely in relation to the initiation of GREI. However, UT Austin-affiliated datasets published prior to 2023, of which there appear to be at least several dozen to several hundred based on a web interface search, still do not have affiliation metadata in DataCite, (e.g., Jayadev et al., 2020; House & Boehm, 2021; Solomon-Lane et al., 2022). The data that can be obtained from the DataCite API will thus underestimate the total volume of affiliated Mendeley Data deposits and give a misleading impression that it was only used by affiliated researchers within the past few years.

Other examples beyond inconsistently recorded affiliation metadata include standardizing how information in fields is entered (e.g., the ability for researchers to use 'First Name Last Name' or 'Last Name, First Name' in Dataverse installations creates marked heterogeneity), enriching records with PIDs like ORCIDs and ROR IDs where possible, and embedding connections between datasets and other scholarly objects in metadata. The latter point, already being explored in various efforts to build knowledge graphs (e.g., OpenAIRE; Manghi et al., 2019), will require investment from scholarly publishers as well. For example, even articles with linked mediated Figshare deposits often do not record the relationship to that dataset in the article's Crossref metadata entry. Similarly, the reverse linkage — articles to datasets — is also important; in current workflows, authors often do not have the final article DOI when they publish a dataset (and may be required to publish the dataset before the manuscript can proceed to production) and are unlikely to return to the dataset simply to update this metadata. Developing automated or manual processes to enhance metadata, and perhaps even data themselves, stands to benefit a wide range of stakeholders, from increasing discoverability and reusability for potential reusers to increasing findability for funding agencies and institutions. The model employed by Dryad, in which an automated process periodically scrapes Crossref for articles and attempts to identify matches for datasets that currently have no linked article (Lowenberg et al., 2019; Habermann, 2023, 2024), is one example of how repositories can support enhanced metadata.

### 4.4.3. Figshare

The third case study relates to Figshare and is presented as a multi-faceted case study that touches not only on metadata plasticity but also on re-curation, the value of data sharing, and a common re-use case (reproducibility) based on re-examination of the RADS dataset (Mohr & Narlock, 2024). Readers may recall that I previously examined which journals and publishers were associated with Figshare datasets that do have affiliation metadata. In order to assess this for the RADS Figshare datasets, I re-queried each Figshare DOI. One of the fields that I retrieved in this process was affiliation metadata (not present in the CSV versions of the RADS data, which is what I originally referenced, but present in the RData versions), and I happened to notice that nearly half of these deposits did not return affiliation metadata matching any one of the six RADS institutions (Table 9); many included no affiliation metadata at all.

Table 9. Summary of whether Figshare datasets previously reported with a RADS institution affiliation still include this linking metadata. *This count includes some duplication when a dataset was previously listed with multiple RADS institutions. These counts represent the filtering of the RADS dataset to remove all Figshare entries ending in '.v*' (i.e. retaining only the 'parent' DOI) but without consolidation based on the linked *relatedIdentifier*. Matching was done using the institutional strings used in the RADS DataCite query. 'Percent matched' is essentially the percent of datasets that "should" have been retrieved, assuming that all current DataCite affiliation metadata is accurate.

| Institution | Total count* | Unmatched count | Percent matched |
|---|---|---|---|
| Cornell | 298 | 136 | 54.3% |
| Duke | 91 | 58 | 36.2% |
| Michigan | 185 | 123 | 33.5% |
| Minnesota | 112 | 56 | 50.0% |
| Virginia Tech | 141 | 129 | 8.51% |
| Washington U | 268 | 224 | 16.4% |
| TOTAL | 1,096 | 726 | 33.7% |

Various hypotheses with resultant predictions can be developed to explain the widespread lack of RADS affiliation metadata among purportedly affiliated Figshare deposits. One hypothesis is that the discrepancies related to a methodological flaw in the RADS workflow, either related to the process of searching for DOIs in DataCite or in processing a DataCite output. If this is the case, we should expect to see affiliation discrepancies at appreciable levels for other repositories. Conversely, another hypothesis posits that this is a Figshare-specific process, potentially related to the automation of dataset creation at scale through publisher partners. If this is the case, we would expect to see the issue primarily restricted to Figshare. To this end, I randomly sampled 4,000 dataset DOIs from the RADS dataset and queried each of them for current affiliation metadata through the DataCite API (the code to do this uses a set seed [*random_state*] and thus the resampling can be independently reproduced, but the dataset is also provided in Gee, 2025b). Of the 315 DOIs that were returned without a match to a RADS institution, more than 90% were Figshare deposits (Table 10). This provided compelling evidence that the discrepancy in affiliation metadata is related to Figshare specifically rather than to some aspect of the dataset retrieval workflow.

Table 10. Comparison of unmatched and total counts of records for repositories with at least one unmatched dataset in a random re-sampling of 4,000 dataset DOIs from the RADS dataset (Mohr & Narlock, 2024). A total of 76 repositories were included in this random sampling process, but only those with at least one unmatched dataset are shown here. The total unmatched percent is out of 4,000, not just those repositories with at least one unmatched dataset.

| Repository | Total count | Unmatched count | Unmatched percent |
|---|---|---|---|
| Figshare | 443 | 290 | 65.4% |
| Harvard Dataverse | 1,832 | 6 | 0.327% |
| Neotoma | 32 | 14 | 43.7% |
| Zenodo | 563 | 5 | 0.884% |
| **TOTAL** | **2,870** | **315** | **7.87%** |

Examination of individual datasets for the four repositories with unmatched records provided further evidence that this is primarily a systemic Figshare issue. For Harvard Dataverse, all but one of the unmatched datasets is listed with 'Duke Kunshan University' as an affiliation, indicating a 'false positive' match resulting from non-exact search capability in *rdatacite*; the one remaining dataset (Teague, 2019) is interesting because it lists 'University of Michigan' (a RADS institution) in the metadata on the landing page, but the DataCite entry lists 'MIT' as the only affiliation. The author of that dataset is presently affiliated with MIT but was not at the time of dataset publication, and the dataset has never been versioned. For Zenodo, all unmatched datasets list of the following three 'false positive' affiliations: Duke Kunshan University, Duke Kinsman University, or West Virginia Institute of Technology. Manual examination of unmatched records for Neotoma, a specialist paleontological database (Williams et al., 2018), revealed that for the 14 unmatched datasets, at least one of the listed authors is properly affiliated with the institution the datasets were linked to in the RADS dataset (Minnesota), but for some reason, the affiliation metadata no longer exists in the DataCite record. The proportion of unmatched datasets against the total subsampled is high (nearly 50%), but the manual examination confirmed that there (1) used to be affiliation metadata; and (2) that affiliation metadata correctly linked to a RADS institution. It may be that metadata was accidentally erased in a platform overhaul, as I mentioned previously for the Digital Porous Media Portal. In contrast to these three, for Figshare, most datasets that still have affiliation metadata in the DataCite record only list Chinese-based institutions; more than a third currently lack any affiliation metadata, and random spot-checking did not identify researchers with a previous affiliation with a RADS institution. I did not identify any datasets with the same 'false positives,' although there are some institutions with many overlapping words (e.g., 'University of Virginia'; 'University of Electronic Science and Technology of China'; 'Northwestern Polytechnical University').

I also examined the RData files of the RADS dataset, which preserve all of the original fields and structure from the API response retrieved in 2022 (the CSV versions are flattened in a way that returns 'NA' for fields with lists, so affiliation fields are all listed as 'NA' in the CSV versions). Manual inspection of select Figshare datasets led to the identification of datasets in which at least one author had an extreme number of affiliations, as many as 68 in one example (Zhang et al., 2021, linked to Washington University in St. Louis), with affiliations sometimes spanning multiple continents and, crucially, including at least one RADS institution. Further examination of these authors' online profiles (e.g., ORCIDs) and the affiliations listed on the related articles did not support these extensive affiliation listings or a previous or current RADS affiliation. Finally, to ensure that this peculiar metadata was not the result of the retrieval process, I downloaded the 2023 DataCite public data file (DataCite, 2024) and manually inspected Figshare entries, which showed the same peculiar metadata patterns.

This multi-step exploration provides strong evidence that the discovered metadata discrepancies are not the result of a methodological flaw in the RADS study but rather are the result of issues with how certain metadata in Figshare were created or managed. The majority of purportedly affiliated Figshare deposits that currently show no affiliation to a RADS institution were last updated between February and March of 2024 (the final RADS dataset was collected in late 2022; Mohr & Narlock, 2024). This is suggestive of a large-scale re-curation of metadata by Figshare, although I am not aware of any public acknowledgement of such an action, and Figshare did not respond to an email query requesting information on this discovery. In some instances, the affiliation metadata appears to have been overwritten (re-curated) to the 'correct' metadata (congruent with that of the linked article; e.g., Zhang et al., 2021), but in many others, the affiliation metadata has been entirely erased (e.g., Li, 2020, linked to the University of Minnesota; Zheng et al., 2022, linked to Duke). Presumably, it was not just RADS institutions' datasets that were impacted by this process, and large-scale deletion of affiliation metadata without replacement could explain why the number of affiliated Figshare deposits that are currently retrievable through the DataCite API is relatively low. Without express confirmation from Figshare, this remains merely an informed hypothesis, but it is difficult to conceive how hundreds of datasets that were purportedly affiliated with just six institutions in 2022 are suddenly no longer affiliated.

This final case study underscores several points. Firstly, it emphasizes the importance of data sharing and of depositing static versions of datasets specific to a study when working with dynamic resources like APIs where data and/or metadata can be deleted or altered in a way that does not preserve (or publicly expose) previous versions. Although provision of code and parameters may be sufficient to repeat a query, if the underlying data have been altered, a failure to reproduce the results could still occur (such code and parameters are provided for the RADS study; Johnston et al., 2024; Mohr & Narlock, 2024). I did not set out to test the reproducibility of the RADS study but was able to do so in this instance thanks to the provision of the data necessary to test my hypothesis. Secondly, this case study underscores the potential for complications arising from automation of some or all aspects of dataset publication. Automation facilitates efficient scaling but also requires rigorous and continual quality control to ensure that an automated process has not deviated from its intended function or introduced unexpected errors. Figshare's process of programmatically creating and publishing datasets is, to the best of my knowledge, unique among generalist repositories, and although it has certain benefits (e.g., synchronized time-release with the associated article, a coveted feature by researchers), this case study demonstrates that it can introduce quality issues at large scales, and previous sections have demonstrated areas in need of improvement (e.g., automated keyword creation). Finally, this case study underscores the dynamic nature of metadata and the value of open infrastructure that provides dynamic, real-time access to those metadata (e.g., public REST APIs). Static public data files are sometimes recommended for large-scale retrieval, with recommendations from both Crossref and DataCite to keep these up to date by using a REST API to retrieve records added after the release of the static file, but as this case study demonstrates, corpuses of metadata can be altered not only by the addition of novel records but also by modification of previously sampled records.

## 4.5. The value of a custom solution

Ideally, this preprint will also have reinforced the value of a custom solution, similar to those that have been developed to varying degrees by other scholars who have undertaken institutional data discovery projects. Although the frequent lack of an existing foundation often requires significant work to explore mechanics of APIs, repository practices, and other attributes, ideally the continued development of workflows like this will reduce the technical and conceptual barriers to undertaking such work. There is a clear importance to examining data and metadata records at a granular level in large corpuses of research datasets, as variation between and within repositories at the repository-scale (e.g., granularity of DOI minting, handling of timestamps), probably not all of it intentional, can otherwise be difficult to detect, especially without specific knowledge of a given repository's current or previous workflows and any discrepancies in between. Many aspects of this workflow were developed and refined by this rather time-intensive exploration to better understand how metadata were created, crosswalked to, or omitted from, the DataCite/Crossref record, and curated, but ideally at least some of the points noted here and in others' work will reduce the need for others to also undertake such journeys. Scholars engaging in this work would also benefit from increased transparency from repositories when major changes are made to metadata processes.

     As has hopefully also been demonstrated, there are an extensive number of edge cases and nuances that were encountered in this process as well. Because this workflow was developed in the context of a specific institution, I was able to investigate edge cases and discrepancies more easily and at a more granular level than would often be possible in a commercial solution that processes data for all research institutions and thus able to rapidly develop targeted fixes to resolve inconsistencies and solutions. I was also able to develop targeted workflows that are relevant for UT Austin (but perhaps only for some, not all, other R1 institutions; e.g., NCBI workflow) or that work around specific repositories' infrastructure and workflows (e.g., how to discover affiliated Figshare deposits without affiliation metadata).

# 5. Future directions

## 5.1. Increased discovery capabilities

This workflow is intended to be continually developed into the future, albeit at a slower pace; the release of this preprint coincides with reaching a point at which I feel that the core process is relatively stable and comprehensive in light of the many external limitations. Beyond the typical maintenance of the codebase and refactoring to make it more efficient and reusable, I do intend to continue to develop workflows to increase the capture potential. For example, developing a secondary workflow to identify mediated Figshare deposits with Crossref DOIs remains an outstanding issue given their estimated volume.

Developing targeted workflows for certain repositories is also planned. For example, UT Austin is an institutional member of the Qualitative Data Repository (QDR; Karcher et al., 2016), but this workflow did not retrieve any QDR datasets. Searches for affiliated datasets in the web interface and subsequent investigation of such datasets revealed that the lack of retrieval is related to granular metadata that include information beyond the institution. Other repositories may be based on my own experience publishing data as a scientist. For example, MorphoBank (O'Leary & Kaufman, 2011; Long-Fox et al., 2024), a repository for phylogenetic character matrices, is my most used data repository and a popular one among phylogeneticists (including others based at UT Austin; e.g., Parker et al., 2022), but it has never crosswalked affiliation metadata. Both of these examples are useful because there is *a priori* knowledge of usage by UT Austin researchers, possibly providing case studies that can be used to test development of workflows to circumvent their metadata shortcomings, which are hardly exclusive to these repositories.

Additionally, we want to develop workflows that are able to target either specific datasets or repositories that may generate data outside of the canonical workflow of direct association with a scholarly article. One example is MorphoSource (Boyer et al., 2016), a specialist repository for non-clinical computed tomographic data that is extensively used by the University of Texas Computed Tomography (UTCT) lab, a national shared multi-user facility. In collaboration with other institutions, the UTCT lab was involved in the NSF-funded openVertebrate (oVert) Thematic Collections Network project (Blackburn et al., 2024), which sought to digitize an extensive number of natural history specimens and which also provided support for MorphoSource. At the time of this preprint, there are nearly 3,500 deposits on MorphoSource that were generated by the UTCT lab; most (<500) do not have DOIs. Some of these have been generated outside of the oVert project and were published in association with a specific article, but many other deposits were not generated for a specific publication and have not been directly used in an article (though they certainly could be). For UT Austin, datasets generated in this fashion are also of interest in broadly understanding the university's research output, even if they may lie outside of the canonical publication route in direct association with a specific article; just as broadening the notion of data reuse is important, so too is broadening the notion of data generation.

## 5.2. Usage of acquired data

Beyond simply discovering and indexing datasets, the UT Libraries are interested in harnessing the data gleaned from this workflow for making improvements to services around research data and software. Firstly, this workflow has already facilitated the identification of metadata discrepancies or issues with the crosswalk to DataCite, leading to implemented or planned changes in UT Austin-managed repositories. We also hope to use this work to identify potential areas for automated re-curation of metadata. For example, now that ROR identifiers are enabled in TDR, I am developing a script that determines which datasets lack the ROR ID and then use a scripted process to update entries with a current non-ROR affiliation entry that is associated with UT Austin ('University of Texas at Austin'). A similar process could be developed for programmatic ORCID re-curation as well, albeit more cautiously given the challenges of human name disambiguation in the absence of PIDs like an ORCID.

I also intend to build out the workflow in ways that will make it more useful to other units within the UT Libraries and other units on campus. For example, one of the other avenues that is intended for development is semi-automation of notifications of new dataset publications for our subject liaison librarians. We already have a workflow that uses the Dataverse API to retrieve information on new datasets in TDR and then automatically creates populated email drafts in Microsoft Outlook with information about recently published datasets to be sent to liaisons on a monthly basis. Expanding this work for all datasets that can be discovered can help give liaisons a more complete picture of their communities' outputs and repository preferences, especially for those who use specialist repositories rather than TDR, and in turn enable liaisons to make more data-driven decisions. This work to develop assessment tools in addition to some of those that have been shown here (e.g., quantifying how many 'datasets' have, or exclusively comprise, software) is intended to broadly improve how research data (and software) services are provided at UT Austin. Identifying both trends and specific datasets (e.g., ones with exemplar README documentation) will be used to inform how the libraries develop various resources (LibGuides, one-off workshops, recurring workshops) and target educational efforts towards researchers.

# 6. Data sharing statement

The data associated with this version of the preprint are deposited in the Texas Data Repository (Gee, 2025b). Data will only be versioned in association with subsequent versions of this preprint. The RADS dataset reanalyzed here is the '*All_dois_20221119.csv*' file included in the Zenodo deposit (Mohr & Narlock, 2024).

# 7. Code sharing statement

The code associated with this version of the preprint is deposited on Zenodo (Gee, 2025a). The code will continue to be developed on GitHub (https://github.com/utlibraries/research-data-discovery), with additional static deposits made in tandem with future versions of the preprint.

## 8. Acknowledgements

Thanks to Michael Shensky for providing early direction for this project and support and feedback throughout this process. Discussions with the many people who have attended a conference presentation on this work and asked questions or made suggestions were also instrumental in its development. The following organizations are thanked for making their APIs and other databases publicly available: Crossref, DataCite, Dataverse, Dryad, Figshare, the National Library of Medicine (NLM), OpenAlex, and Zenodo.

multi-institutional analysis of where research data are shared. *PLOS ONE*, *19*(4), e0302426. https://doi.org/10.1371/journal.pone.0302426

Kambouris, S., Wilkinson, D. P., Smith, E. T., & Fidler, F. (2024). Computationally reproducing results from meta-analyses in ecology and evolutionary biology using shared code and data. *PLOS ONE*, *19*(3), e0300333. https://doi.org/10.1371/journal.pone.0300333

Karcher, S., Kirilova, D., & Weber, N. (2016). Beyond the matrix: Repository services for qualitative data. *IFLA Journal*, *42*(4), 292–302. https://doi.org/10.1177/0340035216672870

Kenyon, J., Sprague, N., & Flathers, E. (2016). The journal article as a means to share data: A content analysis of supplementary materials from two disciplines. *Journal of Librarianship and Scholarly Communication*, *4*, eP2112. https://doi.org/10.7710/2162-3309.2112

Krawczyk, M., & Reuben, E. (2012). (Un)available upon request: Field experiment on researchers' willingness to share supplementary materials. *Accountability in Research*, *19*(3), 175–186. https://doi.org/10.1080/08989621.2012.678688

Lafia, S., & and Kuhn, W. (2018). Spatial discovery of linked research datasets and documents at a spatially enabled research library. *Journal of Map & Geography Libraries*, *14*(1), 21–39. https://doi.org/10.1080/15420353.2018.1478923

Larter, L., & Ryan, M. (2023). *Female preferences for more elaborate signals are an emergent outcome of male chorusing interactions in túngara frogs* [Dataset]. Dryad. https://doi.org/10.5061/dryad.7d7wm37zs

Li, J. (2020). *Visualization1-FALCON.avi* [Media]. Figshare. https://doi.org/10.6084/m9.figshare.11956734.v1

Lichtenberg, E. M., Mendenhall, C. D., & Brosi, B. (2017). *Dataset supplementing Lichtenberg et al. (2017) Foraging traits modulate stingless bee community disassembly under forest loss. Journal of Animal Ecology* [Dataset]. Zenodo. https://doi.org/10.5281/zenodo.843615

Lin, J. (2018, August 12). Peer review publications [Website]. *Crossref Blog*. https://doi.org/10.64000/gp78m-kkk93

Löffler, F., Wesp, V., König-Ries, B., & Klan, F. (2021). Dataset search in biodiversity research: Do metadata in data repositories reflect scholarly information needs? *PLOS ONE*, *16*(3), e0246099. https://doi.org/10.1371/journal.pone.0246099

Long-Fox, B., Andruchow-Colombo, A., Jariwala, S., O'Leary, M., & Berardini, T. (2024). Addressing Global Biodiversity Challenges: Ensuring Long-Term Sustainability of Morphological Data Collection and Reuse through MorphoBank. *Biodiversity Information Science and Standards*, *8*, 529–537. https://doi.org/10.3897/biss.8.135124

Loster, S., & Krzton, A. (2025, March 12). *A multi-strategic approach to locating institutional data deposits*. Research Data Access and Preservation (RDAP) Annual Meeting, Virtual. OSF. https://osf.io/s4yez/

Lowenberg, D. (2021, February 8). Doing it Right: A Better Approach for Software & Data. *Dryad News*. https://blog.datadryad.org/2021/02/08/doing-it-right-a-better-approach-for-software-amp-data/

Lowenberg, D., Habermann, T., & Chodacki, J. (2019, December 1). *Using persistent identifiers to rescue connections between organizations and their work* [Talk]. https://ui.adsabs.harvard.edu/abs/2019AGUFMED21A..01L

Maitner, B., Santos Andrade, P. E., Lei, L., Kass, J., Owens, H. L., Barbosa, G. C. G., Boyle, B., Castorena, M., Enquist, B. J., Feng, X., Park, D. S., Paz, A., Pinilla-Buitrago, G., Merow, C., & Wilson, A. (2024). Code sharing in ecology and evolution increases